\documentclass[twocolumn,secnumarabic,amssymb, nobibnotes, aps, prd]{revtex4-2}
\usepackage{bm}
\usepackage{graphicx}
\usepackage{booktabs}
\usepackage{longtable}

\begin{document}

\title{Analytic imaging formation analysis for Dark Matter halos: geometric ray tracing and caustics using the gravitational refraction law}%

\author{Omar de J. Cabrera Rosas}%
\email{omcbrss@gmail.com}
\author{Tonatiuh Matos}
\affiliation{Departamento de F\'isica, Centro de Investigaci\'on y de Estudios Avanzados del IPN, A.P. 14-740, 07000 M\'exico Distrito Federal, M\'exico}
%\date{}%
%\date{\today}
\begin{abstract}
  One of the most challenging open questions in physics today is discovering the nature of dark matter. In this work we study the imaging formation in dark matter (DM) halos due to an external light source using some DM profiles for comparison with astronomical observations. Approaching these models on a small scale, we analyze the images generated on the lens plane by obtaining the analytical scaled surface mass densities $\Sigma_{*}(x)$ and their corresponding deflection angles $\alpha_{*}(x)$, for later applying a method for ray tracing using the gravitational refraction law. The method is able to locate the positions of the images on the lens plane, by mapping fringes that represent possible sources (such as other galaxies), placed on the source plane. The regions where the strong lensing occurs for each profile, are determined by fixing the $\lambda$ parameter that establishes the ray tracing process. It is shown that the presence of Einstein rings generated by each profile is directly related with the central branch of the caustic. This method gives us a possible alternative way to distinguish between different DM candidates by observing imaging from external sources.
\end{abstract}
\maketitle
%\tableofcontents

\section{Introduction}
Although the discovery of dark matter (DM) has a long history, the main question about it remains open, what is the nature of DM? Throughout this history there have been many candidates to describe its nature. Two of them that remain as favorites are that the DM is dust composed of particles, the so-called Cold Dark Matter (CDM) and that the DM is a field with ultralight mass and zero spin, the Scalar Field Dark Matter (SFDM) model \cite{Matos:1998vk}, also called Fuzzy, Ultralight, etc. DM. Both candidates are able to reproduce cosmological observations very well. However, for galaxy observations, the CDM model needs additional physical phenomena that are sometimes controversial. In this work we focus essentially on these two models.\\
The most important observable of the DM in galaxies is the dynamics of their stars and gas, the rotation curves of the baryonic matter in them. In recent years, this dynamic has been studied with the observation and analysis of the rotation curves of several samples of galaxies. From these observations an acceleration has been measured, and the conclusion is that the presence of DM is necessary to explain their dynamics \cite{McGaugh:radial, Tonatiuh:masdisc}.\\
The restrictions belonging to the central surface density of the halo, determine the value of the quantity $\mu_{DM}=\rho_s r_s$ which seems to be always a constant, and could be an universal invariant, because there are evidences of this restriction in spiral, dwarf irregulars and elliptical galaxies to name some types \cite{Burkert:structdmhalo}-\cite{Burkert:dmstructure}. It is important to observe that $\mu_{DM}$ is constructed as the product of the characteristic length $r_{s}$ and the central density value $\rho_{s}$ of each galaxy, and this is related with the \textit{soliton} region (the core of the galaxy). This is an important zone, because the extra galactic components does not alter this region and the invariant objects present there, are very helpful to understand the complete behaviour of the galaxy \cite{Ana:conseq}. In fact, it has been well established that the total mass of these systems is $M_{DM}(300\textrm{pc})\sim10^{7}M_{\odot}$ with a characteristic size $r_{s}=300\textrm{pc}$ \cite{Strigari:2008ib}, being the SFDM model the only DM model that can naturally explain these observations \cite{Urena-Lopez:2017tob}; that is, the soliton region is perfectly delimited for its analysis, and it is very important for eventual tests that improves the physics interpretation of the data. Therefore, to study the dynamical process for translating this information to the imaging information, this soliton zone must be analyzed, with the corresponding considerations for that region.\\
The Navarro-Frenk-White (NFW) profile for CDM has been exhaustively studied, because it fits properly with observational data \cite{Meneghetti:cluster,Golse:ellip}; however, it presents problems on a small scale which means that there must be another models, that explain the core zone.\\
In recent years, optic tests have proved to be efficient in the understanding of the stellar dynamics, hence it is essential to have tools that allow us to observe this phenomenon, based on the optical information compiled in astronomical data.\\
For the above reasons, a complete optical analysis of this region is in order, using the proper data that comprehends all the physical information encoded in each model, but at the same time with a simplification of the studied equations. Hence, we use the usual gravitational lensing formalism \cite{Schneider:lensingstrong}, that contains all the mathematical restrictions to fully describe such optical processes, completing it with what is called the \textit{gravitational refraction law} (GRL) for ray tracing, that has been used successfully for describing with analytic equations, the imaging formation in the vicinity of a Schwarzschild black hole \cite{Gilb:gravitoron}. The goal of the method is to compute the images on the lens plane, by literally tracing the images projected onto the lens plane, as functions of the positions of the sources, the source plane and the redshift, which is encoded into the critical surface mass density $\Sigma_{cr}$, that appears in the volumetric density $\rho(r)$.\\
This means to translate the optical information obtained from the corresponding images, and link it with the information of the galactic dynamics.\\
This work is organized as follows. In section 2 we set the basic equations for gravitational lensing; the (normalized) surface mass density and the (normalized) deflection angle, are obtained analytically, for each profile, by using a $\lambda$ parameter that encodes the physical relevant information of the system. In section 3 the lens mapping is defined through the $\mathbf{X}$ vector field, which encodes the gravitational refraction ray $\hat{\mathbf{R}}_{G}$. The jacobian of the mapping, the critical and caustic sets are established, along with the conditions for the jacobian that determine the strong lensing regime (the conditions for image multiplicity). In section 4, the general method for the imaging formation process is established, by setting the mapping for the fringes located on the source plane, and later by calculating the equations that draw such images on the lens plane with the ray tracing process generated by the $\mathbf{X}$ field. In section 5 it is applied the method to the studied profiles, approaching some useful cases to illustrate the procedure; in this section the differences between the caustic zones, the Einstein rings and the images generated in each case, are shown.\\
Finally, in section 6 we present the conclusions of this work.

\section{Basic equations for gravitational lensing}
\begin{figure}
  \centering
  \includegraphics[width=80mm]{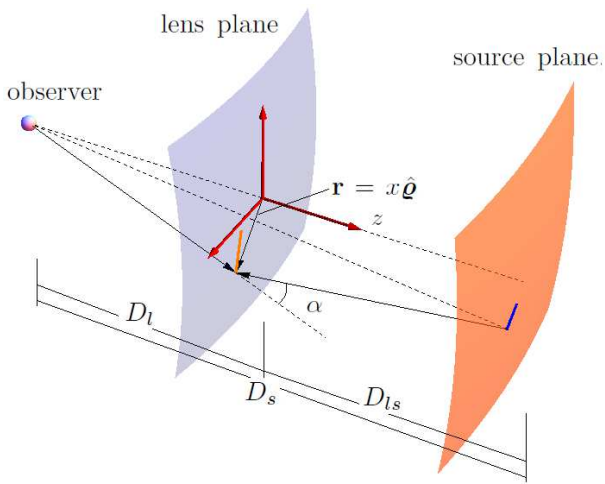}
  \caption{Usual configuration of the lens system. The lens plane vector $\mathbf{r}$, the deflection angle $\alpha$, and the cosmological distances $D_{s}$, $D_{l}$ and $D_{ls}$ are shown. The blue line placed on the source plane represents a linear source (a fringe) that is mapped by the lens mapping to the orange line placed on the lens plane. Also, a cartesian coordinate system whose $z$ axis coincide with the observation axis (in red), is placed on the lens plane to describe the corresponding vector fields of the physical system.}\label{system}
\end{figure}
The configuration for the gravitational lensing system is given in Fig. (\ref{system}), where the source plane, the lens plane and the observer are shown. $D_{s}$, $D_{l}$ and $D_{ls}$ are the distances between observer and source, observer and lens and between lens and source planes, respectively \cite{Schneider:lensingstrong}.\\
Assume that the DM density profile is given in the form $\rho (x)= \rho_{s}f(x)$, with $\rho_{s}$, $r_{s}$ and $x=r/r_{s}$ are the characteristic density, the characteristic radius and the scale radius, respectively; $f(x)$ is a function that gives information about the shape of the profile \cite{Tonatiuh:masdisc}.\\
For finding the deflection angle ($\alpha(x)$) the surface mass density ($\Sigma(x)$) in each case, we use \cite{Schneider:lensingstrong}
\begin{equation}\label{basicLens}
\begin{array}{l}
\displaystyle \Sigma(\xi)=\int_{-\infty}^{\infty}\rho(\xi,z)dz,\\[3ex]
\displaystyle \alpha(x)=\frac{m(x)}{x},
\end{array}
\end{equation}
where the radial coordinate $r$ is related to cylindrical polar coordinates by $r=\sqrt{\xi^{2}+z^{2}}$, and the projected mass $m(x)$ is defined by
\begin{equation}\label{surfacemass}
m(x)=2\pi \int_{0}^{x}\Sigma(\xi)\xi d\xi.
\end{equation}
Using a standard reparametrization \cite{Tonatiuh:strongsfdm, Herrera:strong}, the previous equations can be written in a dimensionless form by defining
\begin{equation}\label{Lambdadef}
\lambda=\frac{\mu_{DM}}{\pi \Sigma_{cr}}=10^{-3}\frac{0.57}{h}\left( \frac{\mu_{DM}}{M_{\odot}\textrm{pc}^{-2}}\right)\frac{d_{l}d_{ls}}{d_{s}},
\end{equation}
where $\Sigma_{cr}=c^{2}D_{s}/(4\pi G D_{l}D_{ls})$ is the critical surface mass density, $\mu_{DM}=\rho_{s}r_{s}$ and $d_{a}=D_{a}H_{0}/c$, are the reduced (dimensionless) angular distances, with $H_{0}$ the Hubble parameter.\\
Eqs. (\ref{basicLens}) and (\ref{surfacemass}) are now given by \cite{Herrera:strong}
\begin{equation}\label{Repara}
\begin{array}{cccc}
\displaystyle \Sigma_{*}(x)=\frac{\Sigma(x)}{\mu_{DM}},& {} & {} & \displaystyle m_{*}(x)=\frac{m(x)}{\rho_{s}r_{s}^{3}} ,\\[2ex]
\displaystyle \alpha_{*}(x)=\frac{m_{*}(x)}{x}, & {} & {} & \displaystyle \lambda_{cr}^{-1}=\pi \Sigma_{*}(0),
\end{array}
\end{equation}
where  it is defined $\lambda_{cr}$, which is the smallest value of $\lambda$ in Eq. (\ref{Lambdadef}) for which an Einstein ring appears \cite{Herrera:strong}.\\
With the above definitions it is possible to write the (scalar) dimensionless lens equation as
\begin{equation}\label{dimlesslens}
\beta_{*}(x)=x-\lambda\frac{m_{*}(x)}{x}.
\end{equation}
%in such a way that the physical parameters of the system are encoded into $\lambda$ (Eq. \ref{Lambdadef}).\\
For $\beta_{*}(x)=0$, the solutions for Eq. (\ref{dimlesslens}) give the radius of the Einstein rings produced by each lens:
\begin{equation}\label{EinRingBeta}
  \lambda_{E}=\frac{x^{2}}{m_{*}(x)}.
\end{equation}
It is important to remark the types of models approached here: the Navarro-Frenk-White profile is obtained as a fit function of the numerical simulations of the CDM model \cite{Navarro:NFW}; for the SFDM we will study the Multistate Scalar Field Dark Matter (MSFDM) model, which is a result of analytical solutions of the Klein-Gordon equations \cite{Robles:2012kt}. The Bose-Einstein condensate (BEC), which is a solution of the Klein-Gordon equations with the Thomas-Fermi approximation \cite{Bohmer:canbec}.  And finally, the Wave DM model, which is similar to the NFW profile but now for the SFDM model \cite{Schive:cosmicdwave}.\\
All of them (except the NFW profile), satisfy the restriction where $\mu_{DM}=\rho_{s}r_{s}$ is the density of the nucleus of the galaxy \cite{Tonatiuh:masdisc, Ana:conseq}.\\
Using Eqs. (\ref{basicLens})-(\ref{Repara}), it is obtained table \ref{TAB}, for these DM profiles (see appendix \ref{AppSur}). \\

\begin{table*}[t]
 {\setlength\doublerulesep{0.4pt}
\begin{tabular}{l c c c c c c c c c c c }
\toprule[1pt]\midrule[0.3pt]
  % after \\: \hline or \cline{col1-col2} \cline{col3-col4} ...
  Name & $f(x)$ & {} & $\Sigma_{*}(x)$ & {} & $\alpha_{*}(x)$ & {} & $\mu_{DM}[M_{\odot}\textrm{pc}^{-2}]$ & {} & $\lambda_{cr}$ & {} & \\[.8ex]
  \small{(1)} & {\small{(2)}} & {} & {\small{(3)}} & {} & {\small{(4)}} & {} & {\small{(5)}} & {} & {\small{(6)}} \\[1ex]\hline\\[.5ex]
  \textbf{BEC} & $\displaystyle \frac{\sin x}{x}$ & {} & $\displaystyle \pi J_{0}(x)$ & {} & $\displaystyle 2\pi^{2} J_{1}(x)$ & {} & 139 & {} & $\frac{1}{\pi^{2}}\approx 0.101$\\[3ex]
  \textbf{MSFDM} & $\displaystyle \frac{\sin^{2}x}{x^{2}}$ & {} & $\displaystyle F(x)_{MS}$ & {} & $\displaystyle \frac{2}{x}g(x)_{MS}$ & {} & 139 & {} & $\frac{1}{\pi^{2}}\approx 0.101$ \\[3ex]
  \textbf{Wave DM} & $\displaystyle\frac{1}{(1+x^{2})^{8}}$ & {} & $\displaystyle \left(\frac{13!!}{2^{7}7!}\right)\frac{\pi}{(x^{2}+1)^{15/2}}$ & {} & $\displaystyle \frac{\pi^{2}}{x}\left(\frac{11!!}{2^{6}7!}\right)\left( 1-\frac{1}{(x^{2}+1)^{13/2}}\right)$ & {} & 648  & {} & $\frac{2^{7}7!}{13!!\pi^{2}}\approx 0.484$ \\[3ex]
  \textbf{NFW} &$\displaystyle\frac{1}{x(1+x)^{2}}$ & {} & $\displaystyle 2F(x)_{NFW}$ & {} & $\displaystyle \frac{4}{x^{2}}g(x)_{NFW}$ & {} & 89  & {} & 0 \\\midrule[0.3pt]\bottomrule[.6pt]
  \end{tabular}}
  \caption{Functions for computing the imaging formation process for DM halos. Second column: $f(x)$ functions corresponding to the volumetric density for each profile. Third and fourth columns: analytical surface mass densities and deflection angles (both normalized, see Appendix \ref{AppSur}) for each DM profile. Fifth column: $\mu_{DM}$ values \cite{Ana:conseq}; it is important to point out that $\mu_{DM}$ (the product of the corresponding values of $r_{s}$ and $\rho_{s}$), is a constant for all cases \cite{Ana:conseq}. Sixth column: values of $\lambda_{cr}$ obtained directly from the condition $\lambda_{cr}=(\pi \Sigma_{*}(0))^{-1}$.}\label{TAB}
\end{table*}

\begin{figure}
  \centering
  \begin{tabular}{l}
  \includegraphics[width=86mm]{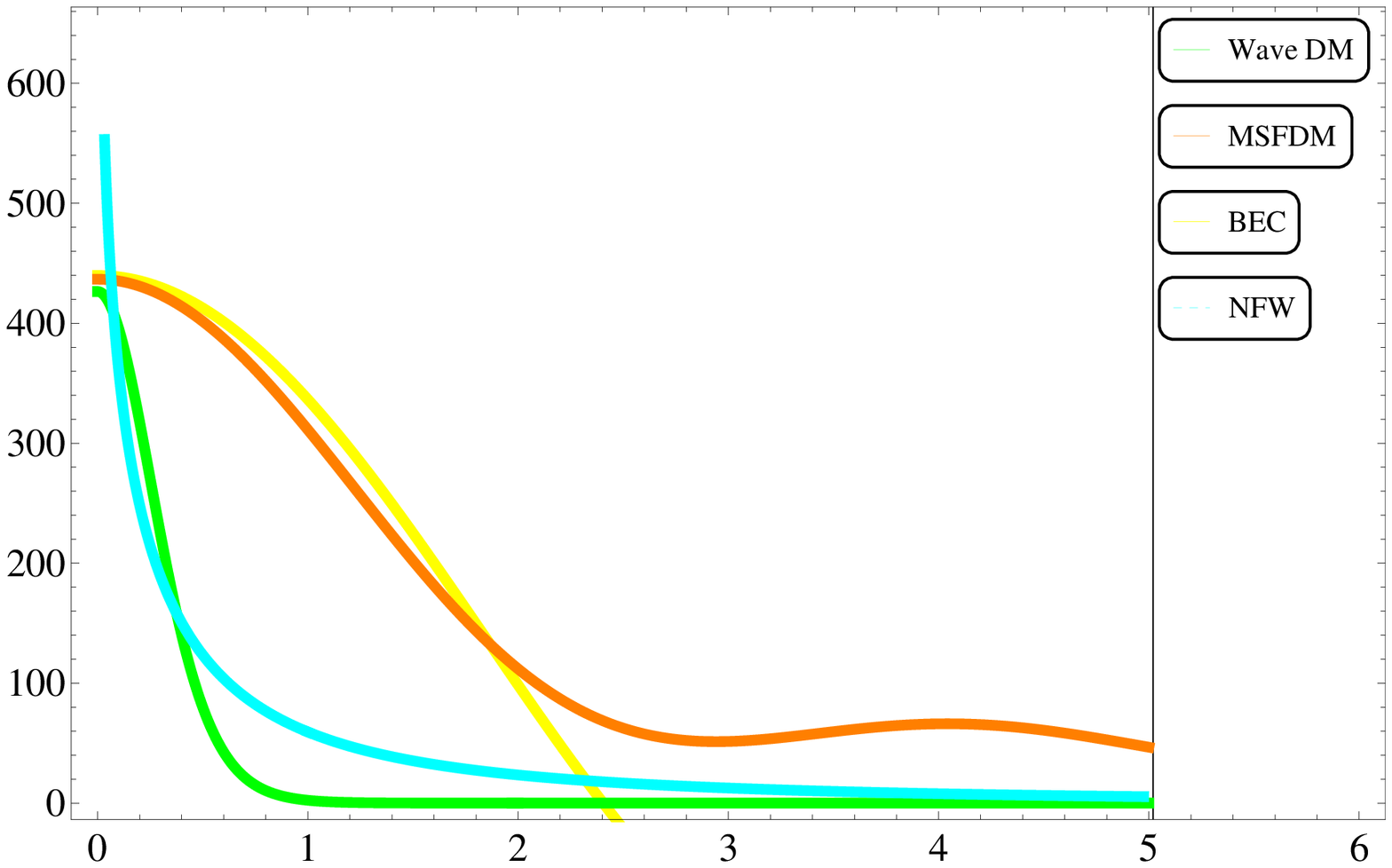}\\[0.5ex]
  \small (a)\\[2ex]
  \!\!\!\!\!\!\!\!\!\!\!\includegraphics[width=95mm]{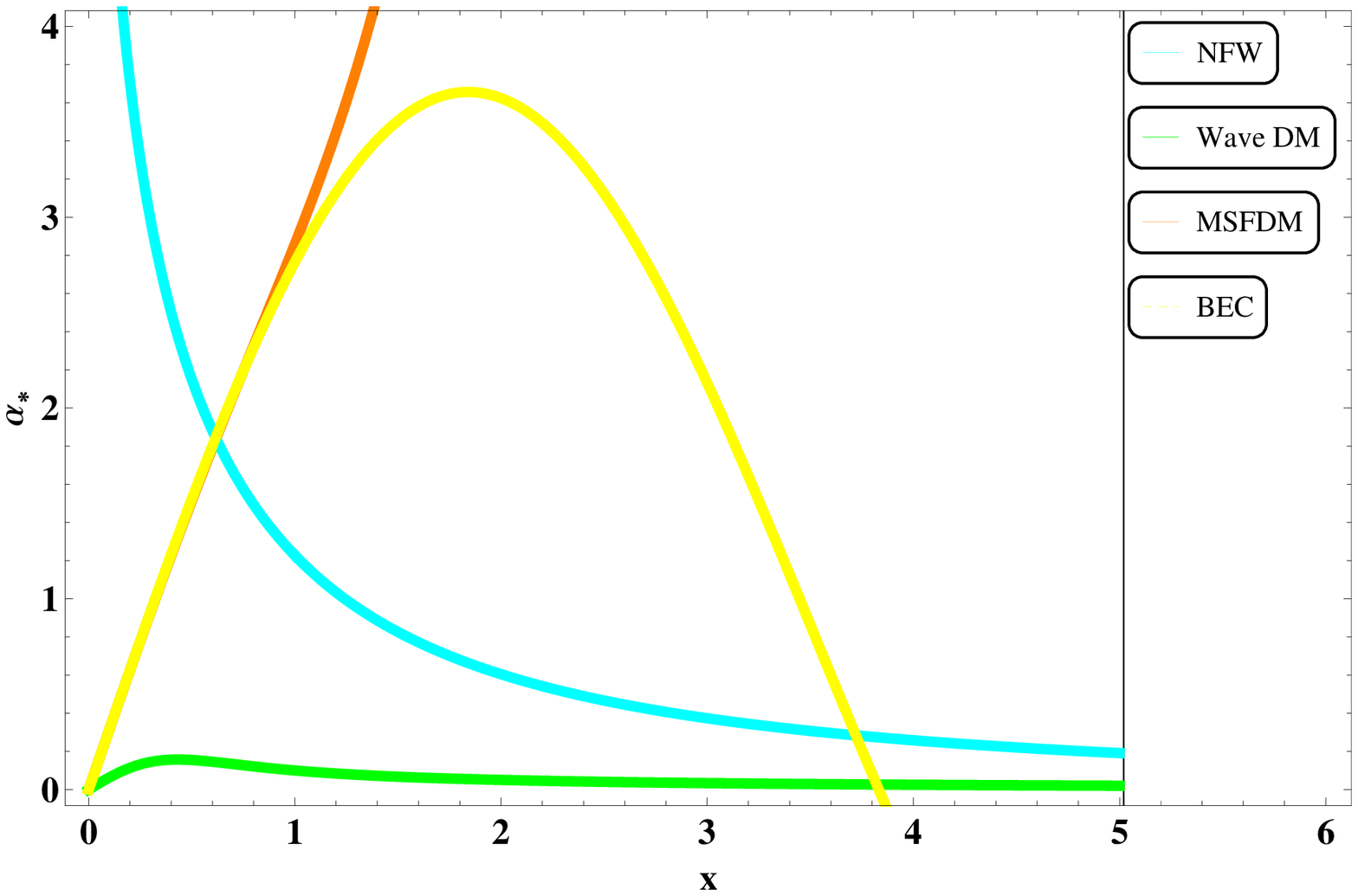}\\[0.5ex]
  \small (b)
  \end{tabular}
  \caption{(a)Comparison of the $\Sigma(x)$ functions for each profile, using the corresponding values of $\mu_{DM}$ obtained from table \ref{TAB}. (b) Plots corresponding to the $\alpha(x)_{*}$ (normalized) angles. Except for the BEC and MSFDM profiles, the image of these functions is always positive. This condition has consequences for the critical set and the region for strong lensing in each case (see Eq. \ref{strongjacob}).}\label{SigmaAlpha}
\end{figure}

\section{Rays and caustics}
\subsection{Gravitational Refraction Law}
To develop the imaging formation process it is necessary to know the deflection angle for each case.\\
For this purpose, it will be helpful to express the quantities for describing the optical process, in a cylindrical coordinate basis $\{\hat{\bm{\varrho}},\hat{\bm{\phi}},\hat{\bm{z}}\}$ (see the red coordinate system from Fig. \ref{system}). To compute the evolution of the light rays, the observer is replaced by a fictitious point light source, for later looking for the points on the lens plane connected via a (fictitious) refracted light ray with points of the source plane (in analogy to the usual refraction process), in such a way that a point $\mathbf{X}$ on a deflected light ray is given by \cite{Gilb:gravitoron}
\begin{equation}\label{Xfield}
\mathbf{X}=\mathbf{r}+l\hat{\mathbf{R}}_{G},
\end{equation}
where $\mathbf{r}=x\hat{\bm{\varrho}}$ is the lens plane vector, $l$ is a distance along the deflected light ray and $\hat{\mathbf{R}}_{G}$ gives the direction of evolution of such light ray.\\
The $\hat{\mathbf{R}}_{G}$ vector can be written as a linear combination of an emission unitary vector $\hat{\mathbf{I}}$ and the $\hat{\bm{\varrho}}$ vector on the lens plane as \cite{Gilb:gravitoron}
\begin{equation}\label{GRLaw}
  \hat{\mathbf{R}}_{G}= \left[ \cos\alpha + \frac{(\hat{\mathbf{I}}\cdot\hat{\bm{\varrho}})\sin\alpha}{\sqrt{1-(\hat{\mathbf{I}}\cdot\hat{\bm{\varrho}})^{2}}}\right]\hat{\mathbf{I}}-\left[ \frac{\sin\alpha}{\sqrt{1-(\hat{\mathbf{I}}\cdot\hat{\bm{\varrho}})^{2}}} \right]\hat{\bm{\varrho}}.
\end{equation}
Eq. (\ref{Xfield}) together with Eq. (\ref{GRLaw}) represent the gravitational refraction law (GRL), analogous to the usual refraction law.\\
Now, under the assumption that the source is placed at infinity ($\hat{\mathbf{I}}=-\hat{\bm{z}}$), Eq. (\ref{GRLaw}) is reduced to
\begin{equation}\label{refract}
  \hat{\mathbf{R}}_{G}=\cos\alpha\hat{\bm{z}}-\sin\alpha\hat{\bm{\varrho}},
\end{equation}
and Eq.(\ref{Xfield}) can be written as
\begin{equation}\label{XangleF}
\mathbf{X}=(\varrho-\l\sin\alpha)\hat{\bm{\varrho}}+\l\cos\alpha \hat{\bm{z}}.
\end{equation}
Now, reparameterizing Eq. (\ref{XangleF}) using Eqs. (\ref{Repara}), for an arbitrary deflection angle $\alpha_{*}(x)$ (in the small angles regime where $\sin\alpha\approx\alpha$ and $\cos\alpha\approx 1$), $l=\lambda$ and with lens plane vector $\mathbf{r}=x\hat{\bm{\varrho}}$, Eq. (\ref{XangleF}) can be rewritten as
\begin{equation}\label{Xreal}
\mathbf{X}=(x-\lambda\alpha_{*})\hat{\bm{\varrho}}+\lambda\hat{\bm{z}}.
\end{equation}
Eq. (\ref{Xreal}) is the principal equation for describing the gravitational optical process (by means of the GRL through the information encoded by the $\mathbf{X}$ field), because encodes all the relevant (geometrical) information that allows to analyze the imaging formation, rays and caustics, analytically.\\
Observe that the lens equation is a map between points on the lens plane ($L$) to the source plane ($S$) \cite{Schneider:lensingstrong}, in the same way that the $\mathbf{X}$ vector field (Eq. \ref{Xreal}) maps points from $L$ to $S$, so this vector field is in fact another representation of the usual lens equation expressed as a three dimensional object. In fact, observe that the $\varrho$ component of the $\mathbf{X}$ vector field ($\hat{\bm{\varrho}}\cdot \mathbf{X}$), is the scalar lens equation (\ref{dimlesslens}):
\begin{equation}\label{componentZ}
  \beta_{*}(x)=\hat{\bm{\varrho}}\cdot \mathbf{X}.
\end{equation}
The above implies that the minimum approach distance along the observation axis, is determined by $\lambda_{cr}$ which is obtained analytically from the $\Sigma_{*}(x)$ functions in table \ref{TAB}.\\
Therefore, we shall use Eq. (\ref{Xreal}) to generate the corresponding images because it contains relevant physical information of the system, and such information will be useful to generate and understand the images that appear on the lens plane (this, together with the information provided by the caustics associated with them).

\subsection{Critical and caustic sets}
The refracted light rays given by an arbitrary deflection angle $\alpha_{*}$ are locally described by Eq. (\ref{Xreal}), which is a mapping between local coordinates in the domain space (the lens plane) $x$, $\phi$ and $\lambda$, and the local coordinates in the target space (the source plane) $X$, $Y$ and $Z$, that is, Eq. (\ref{Xreal}) describes a map from $\mathcal{R}^{3}$ to $\mathcal{R}^{3}$ where the critical set of this map is given by the of points in the domain space such that the map is not locally one to one \cite{Arnld:singdffmps,Arnld:ctstheo}; hence, the critical set is obtained by the condition
\begin{equation}
  \mathcal{J}=\frac{\partial \mathbf{X}}{\partial x}\cdot\left(\frac{\partial\mathbf{X}}{\partial \phi}\times \frac{\partial \mathbf{X}}{\partial \lambda}\right)=0.
\end{equation}
In this case, the jacobian of the system is equivalent to
\begin{equation}\label{jacobparameter}
  \left( 1-\lambda\frac{\alpha_{*}}{x}\right)(1-\lambda\alpha_{*}')=0,
\end{equation}
where the prime denotes the derivative respect to $x$.\\
This condition for the jacobian is equivalent to the condition for finding the \textit{critical curves} of the usual lens mapping $\bm{\mathcal{A}}$, on which the determinant of the $\bm{\mathcal{A}}$ matrix is equal to zero \cite{Schneider:lensingstrong}\\
The solutions of Eq. (\ref{jacobparameter}) are
\begin{equation}\label{criticB}
\begin{array}{lcccl}
\displaystyle \lambda_{+}=\frac{x}{\alpha_{*}},& {} & {} & {} &\displaystyle \lambda_{-}=\frac{1}{\alpha_{*}'},
\end{array}
\end{equation}
and these equations represent the critical set of the mapping.\\
On the other hand, the image of the critical set is the caustic set \cite{Gilb:gravitoron,Arnld:singdffmps}. Evaluating Eqs. (\ref{criticB}) into Eq. (\ref{Xreal}), the caustic set branches are given by
\begin{equation}\label{causticBr}
\begin{array}{l}
\displaystyle \mathbf{X}_{+}= \frac{x}{\alpha_{*}}\hat{\bm{z}},\\[3ex]
\displaystyle \mathbf{X}_{-}= (x-\frac{\alpha_{*}}{\alpha_{*}'})\bm{\hat{\varrho}}+\frac{1}{\alpha_{*}'}\hat{\bm{z}}.
\end{array}
\end{equation}
Due to the symmetries of the system, it can be seen that the $\mathbf{X}_{+}$ branch is a segment of line along the $z$ axis (the observation axis), meanwhile the $\mathbf{X}_{-}$ branch is a surface of revolution along the same axis. The projections on the lens plane of these branches are the \textsl{radial critical curves} (for $\mathbf{X}_{+}$) and the \textsl{tangential critical curves} (for $\mathbf{X}_{-}$) \cite{Schneider:lensingstrong}.\\
Although the $\mathbf{X}_{+}$ branch is always real for all the studied cases, this branch not always evolve in the positive $z$ direction, besides the fact that for some profiles, there is a zone of virtual caustic for the $\mathbf{X}_{-}$ branch, determined by the sign of $\alpha'_{*}$ (see plots \ref{virtualzone}).\\
From Eq. (\ref{Xreal}) it can be seen that $\lambda$ parameterizes the $\hat{z}$ component of the $\mathbf{X}$ vector field, and the minimum value reached by this field along the $z$ axis, is in fact given by $\lambda_{cr}$, in the limit for $x\rightarrow 0$ (the perfect alignment condition \cite{Herrera:strong}), but it is not limited to this situation, because $\lambda_{cr}$ bounds both the critical set and the beginning of the central branch of the caustic $\mathbf{X}_{+}$ (Eqs. \ref{causticBr}). \\
In fact, notice that both $\lambda_{+}$ and $|\mathbf{X}_{+}|$ equate the condition for Einstein rings of the system (Eq. \ref{EinRingBeta}), \textit{i.e.}, $\lambda_{+}=|\mathbf{X}_{+}|=\lambda_{E}$ and this means that such rings evolve with these functions. Therefore, with the interpretation given by the $\mathbf{X}$ field, it means that the Einstein rings for each system, are only produced when the source plane is in contact with the central branch of the caustic; if this branch does not reach a contact point with that plane, there is no formation of such rings (see plots \ref{yRoots} and table \ref{3Dplots}).

\subsection{Conditions for image multiplicity}
\begin{figure*}[t]
\begin{tabular}{c c}
  % after \\: \hline or \cline{col1-col2} \cline{col3-col4} ...
  \includegraphics[width=55mm]{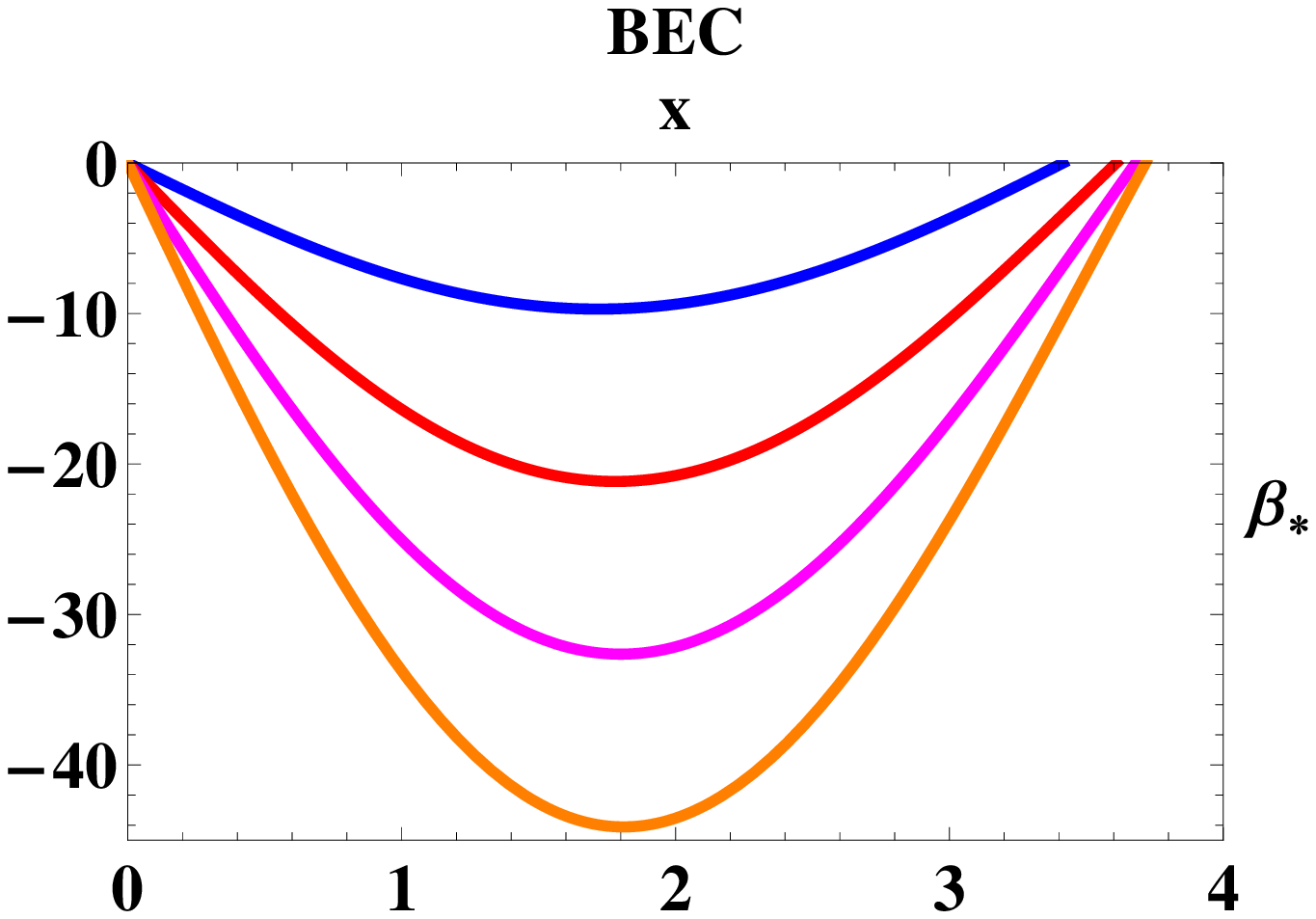} & \includegraphics[width=48mm]{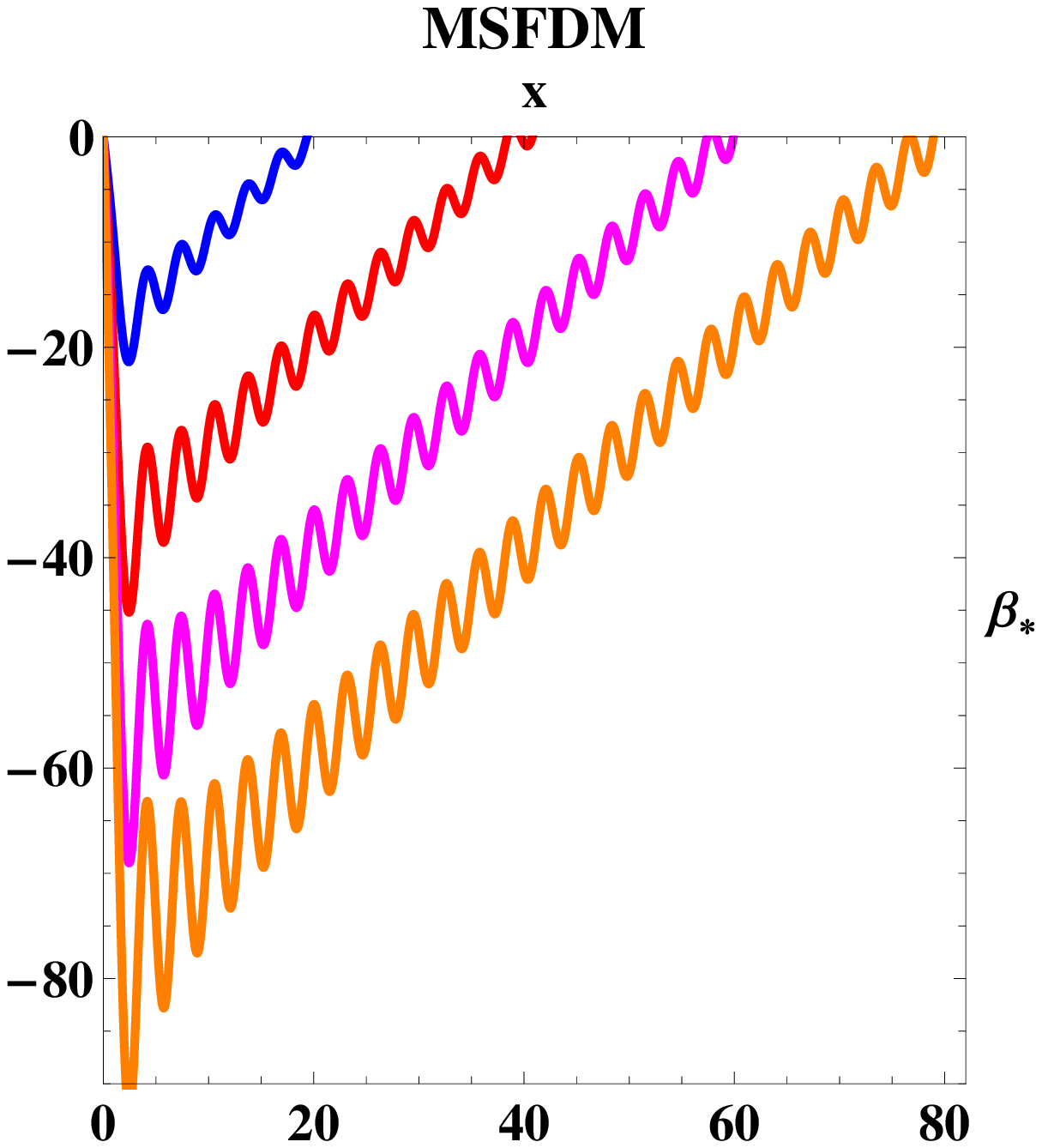} \\[0.5 ex]
  \small (a) & \small (b) \\[0.5ex]
   \includegraphics[width=55mm]{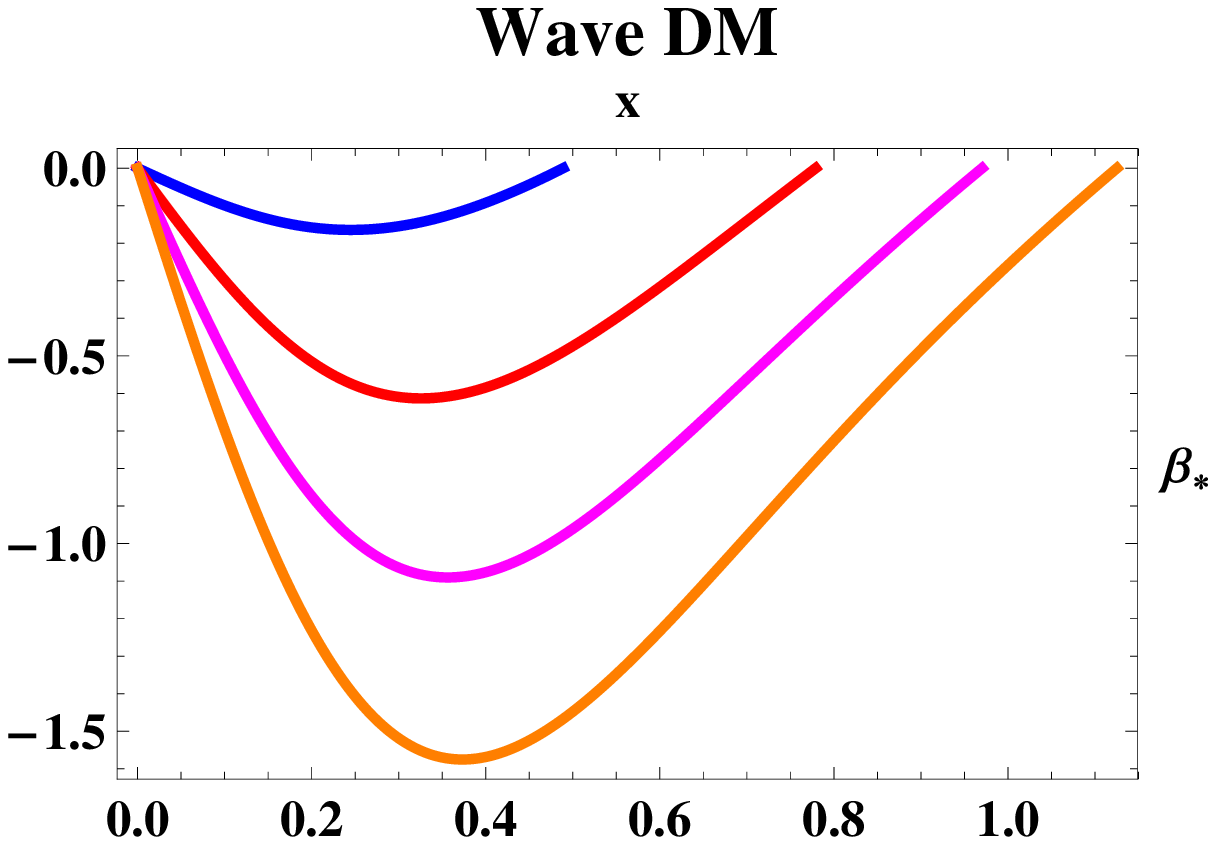}& \includegraphics[width=52mm]{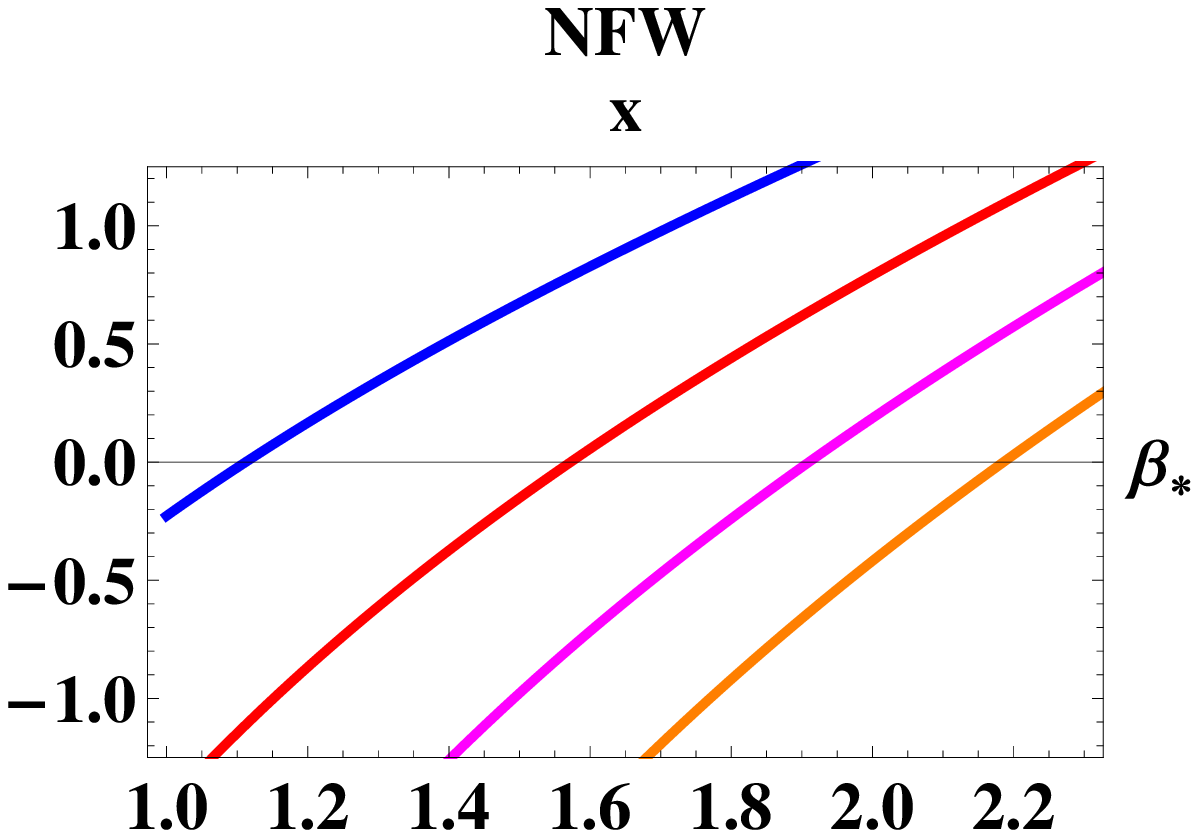} \\[0.5ex]
  \small (c) & \small (d) \\[0.5ex]
\end{tabular}
\caption{Plots of the mapping $y=x-z_{0}\alpha_{*}$ (the lens mapping $\beta_{*}(x)$) for $z_{0}=1 \textrm{(blue)},2 \textrm{(red)},3 \textrm{(magenta)},4 \textrm{(orange)}$. The solutions for $\beta_{*}=0$ denote the possible Einstein rings produced in each case. Notice that because in the Burkert profile, the curves do not intersect the horizontal axis, this profile doesn't have Einstein rings.}
\label{yRoots}
\end{figure*}

The determinant of the $\bm{\mathcal{A}}$ matrix (Eq. \ref{jacobparameter}) defines the region where the lens produces multiple images, if it happens that $\det\bm{\mathcal{A}}<0$ \cite{Schneider:lensingstrong}, which is written simply as
\begin{equation}\label{strongjacob}
  \mathcal{J}=\left( 1-\lambda\frac{\alpha_{*}}{x}\right)(1-\lambda\alpha_{*}')<0.
\end{equation}
By using this condition, the region for strong lensing in each case is determined (see table \ref{strongLENS}), by fixing the values of the $\lambda$ parameter.\\
Another requirement for image multiplicity is that the shear function $\kappa=\Sigma/\Sigma_{cr}$, fulfills that $\kappa>1$ \cite{Schneider:lensingstrong}; this condition can be restated using the $\lambda$ parameter (in Eqs. \ref{Repara}), where $\kappa$ is now written as $\kappa=\pi\lambda\Sigma_{*}=\lambda/\lambda_{cr}$, which implies that $\lambda\geq\lambda_{cr}$ and this determines the minimum value for the (structural)
surface density $\rho_{s}r_{s}$ of any given DM profile in terms of the measured quantities of a lens system \cite {Herrera:strong}.  In this case, this means that $\lambda$ must be greater (or at least equal) than the minimal value of the central branch of the caustic along the observation axis (see tables \ref{TAB} and \ref{centralXbranch}), $\lambda_{cr}$.  Recalling now that the caustic also contributes to the images present on the lens plane (a change to the source position does not lead to the change of the number of images unless a source moves along a caustic branch \cite{Schneider:lensingstrong}), we study then the behaviour of that central branch, because when is real (not virtual), it is possible that the source plane has at least one point in contact with it, depending on the conditions imposed geometrically by the $\Sigma_{*}(x)$ function.\\
Now, a reparametrization of Eq. (\ref{XangleF}) can be found by taking $z_{0}=l\cos\alpha$, and with the restrictions for small angles, this implies that $z_{0}=\lambda$, and hence if the source plane is placed at $z=z_{0}$ and all the values of $x_{0}$ such that $y(x_{0},z_{0})= x_{0}-z_{0}\alpha(x_{0})=0$ are found, the roots of this mapping represent the values of $x$ where a circle appears; this circle is the Einstein ring corresponding to different positions of the source plane $z_{0}$ \cite{Gilb:gravitoron}. Therefore, this last condition is completely equivalent to Eq. (\ref{EinRingBeta}).\\
In the plots from Figs. (\ref{yRoots}) the roots for the $y=\beta_{*}$ mapping are shown for each profile and although $z_{0}$ can take continuous values, we show only the roots for discrete values of $z_{0}=1 \textrm{ (blue)},2 \textrm{ (red)},3 \textrm{ (magenta)},4 \textrm{ (orange)}$.\\

\section{Imaging Formation}
Consider a family of one dimensional sources in the region $z>0$, locally described by
\begin{equation}\label{sigmalabel}
\mathbf{X}=\mathbf{X}(\sigma,n),
\end{equation}
where $n$ denotes the source and $\sigma$ labels the points on that source. From Eqs. (\ref{Xreal}) and (\ref{sigmalabel}), the images that the observer sees on the lens plane $z=0$, are given by all the points $(x\cos\phi,x\sin\phi,0)$, where $x$ and $\phi$ are solutions to \cite{Gilb:gravitoron}
\begin{equation}\label{planemap}
\begin{array}{l}
\displaystyle X_{s}(\sigma,n)= [x-Z_{s}(\sigma,n)\alpha(x)]\cos\phi,\\[2ex]
\displaystyle Y_{s}(\sigma,n)= [x-Z_{s}(\sigma,n)\alpha(x)]\sin\phi.\\[2ex]
\end{array}
\end{equation}
If the sources (of length $L_{b}-L_{a}$) make an angle $\Phi$ with the positive \textsl{x} axis on the plane $z=z_{0}$, then
\begin{equation}\label{inclined}
\begin{array}{c}
\displaystyle Y_{s} = X_{s}\tan\Phi+n,\\[2ex]
\displaystyle Z_{s}=z_{0},
\end{array}
\end{equation}
with $n$ a constant.\\
The equations for obtaining the image for each fringe are determined by all the points $(x,\phi)$ in the lens plane such that \cite{Gilb:gravitoron}
\begin{equation}\label{Ronchifringe}
\begin{array}{c}
X_{s}=[x-z_{0}\tan\alpha(x)]\cos\phi,\\[2ex]
X_{s}\tan\Phi + n=[x-z_{0}\tan\alpha(x)]\sin\phi.
\end{array}
\end{equation}
From these last equations, it can be shown that a parametric analytical expression for the fringes images on the plane $z=0$ is given by \cite{Gilb:gravitoron}
\begin{equation}\label{ronchi}
\begin{array}{l}
\displaystyle T_{x}(x)=x\cos\left[ \Phi + \arcsin\left( \frac{n\cos\Phi}{x-z_{0}\tan\alpha(x)}\right) \right],\\[2ex]
\displaystyle T_{y}(x)=x\sin\left[ \Phi + \arcsin\left( \frac{n\cos\Phi}{x-z_{0}\tan\alpha(x)}\right) \right],
\end{array}
\end{equation}
where $T_{x,y}$ are the coordinates of the images on the lens plane. These equations represent the \textsl{gravitoronchigram}\cite{Gilb:gravitoron}.\\
In our case of study, we are not interested in obtaining the full ronchigram for each profile (\textsl{i.e}, for several values of $n$), but instead we analyze the image of a fringe on a fixed value of $n=n_{0}$, with its length parameterized by $X_{s}$. For simplicity, we take the case of a line source that coincide with the \textsl{x} axis ($\Phi=0$) that represents a linear source, so Eqs. (\ref{Ronchifringe}) are reduced as
\begin{eqnarray}
% \nonumber % Remove numbering (before each equation)
  {} & X_{s}=[x-z_{0}\alpha_{*}(x)]\cos\phi,\label{Xsurface}\\
  {} & n=[x-z_{0}\alpha_{*}(x)]\sin\phi, \label{nsurface}
\end{eqnarray}
with $L_{a}\leq X_{s}\leq L_{b}$.
Due to the symmetries of the system, these equations fulfill the constraints
\begin{eqnarray}
  {} & \displaystyle \phi=\arctan\left( \frac{n}{X_{s}}\right),\label{Constraint1}\\
  {} & X_{s}^{2}+n^{2}=[x-z_{0}\alpha_{*}(x)]^{2}. \label{Constraint2}
\end{eqnarray}
\begin{figure*}[t]
\begin{tabular}{c c}
  % after \\: \hline or \cline{col1-col2} \cline{col3-col4} ...
  \includegraphics[height=50mm,width=55mm]{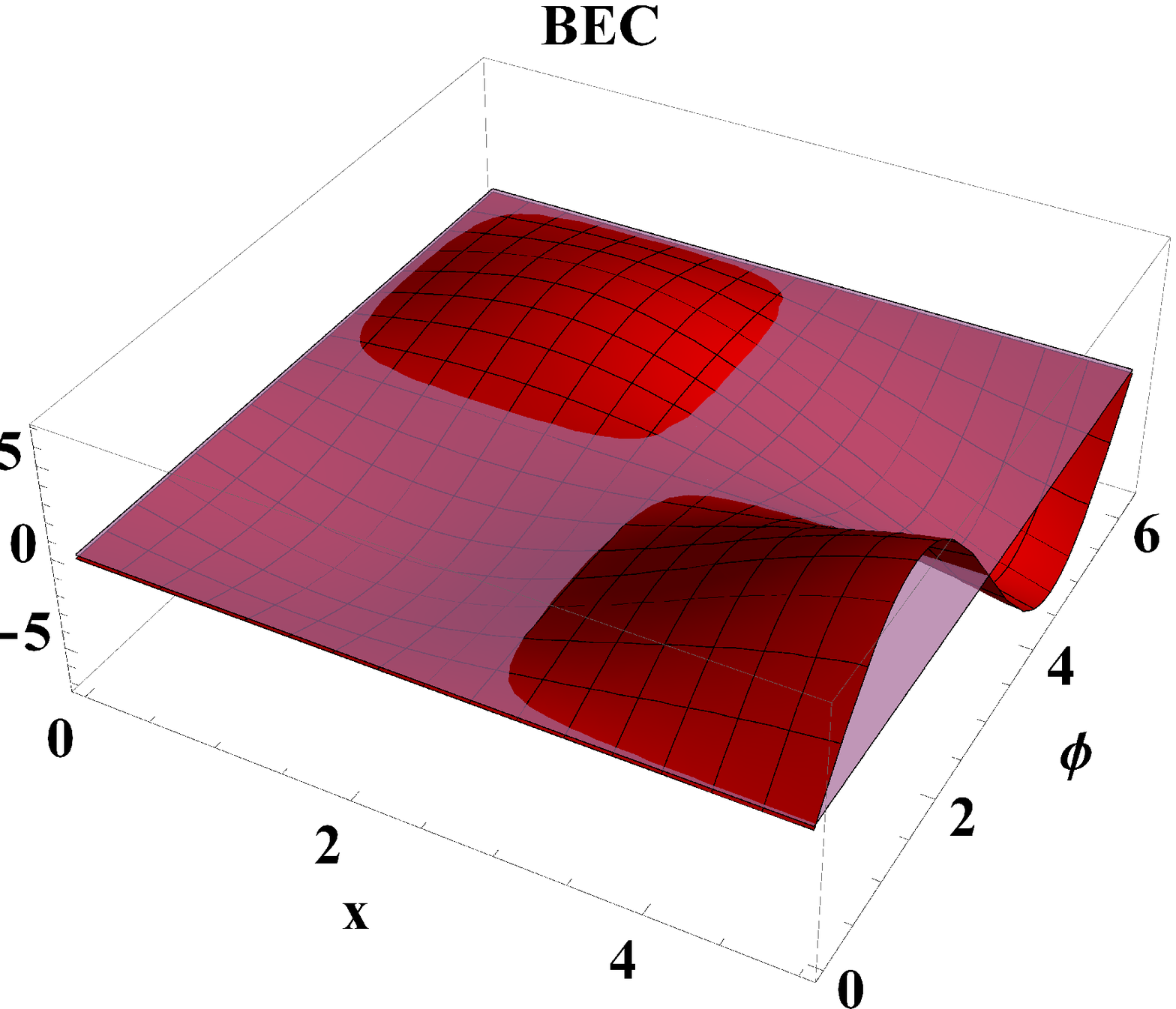}  & \includegraphics[height=50mm,width=55mm]{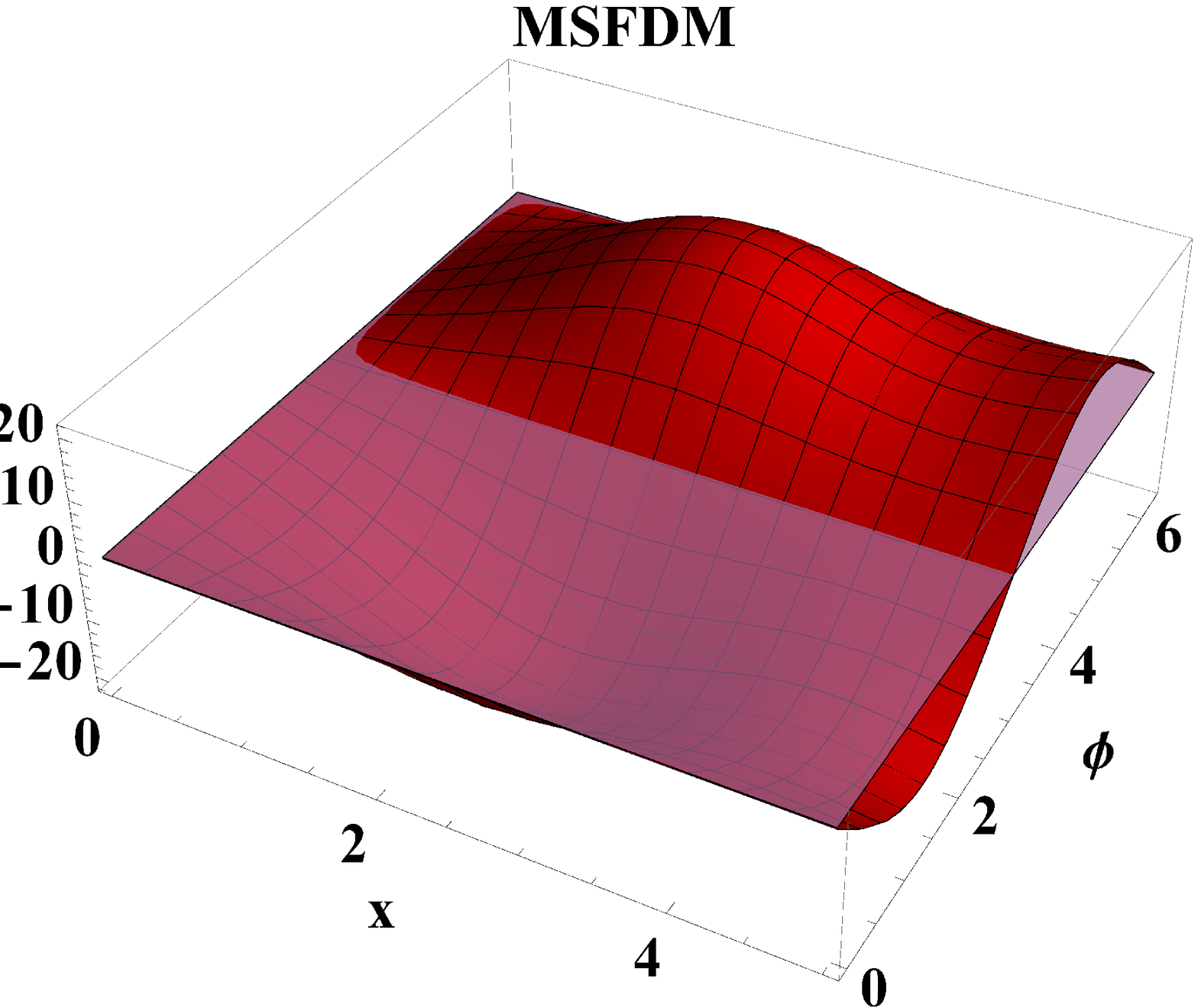} \\[0.5ex]
  \small (a) & \small (b)\\[0.5ex]
  \includegraphics[height=50mm,width=53mm]{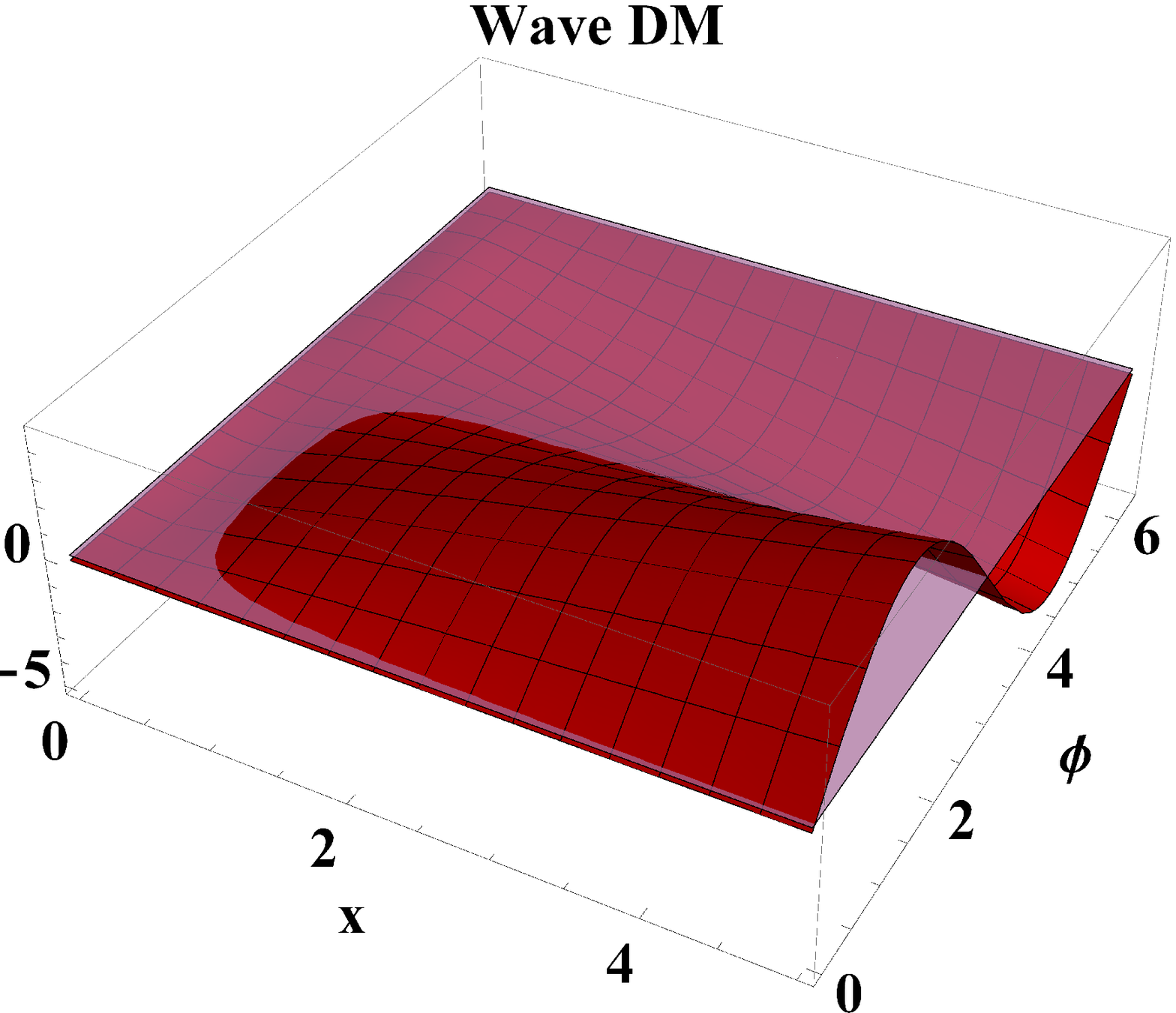} & \includegraphics[height=55mm,width=55mm]{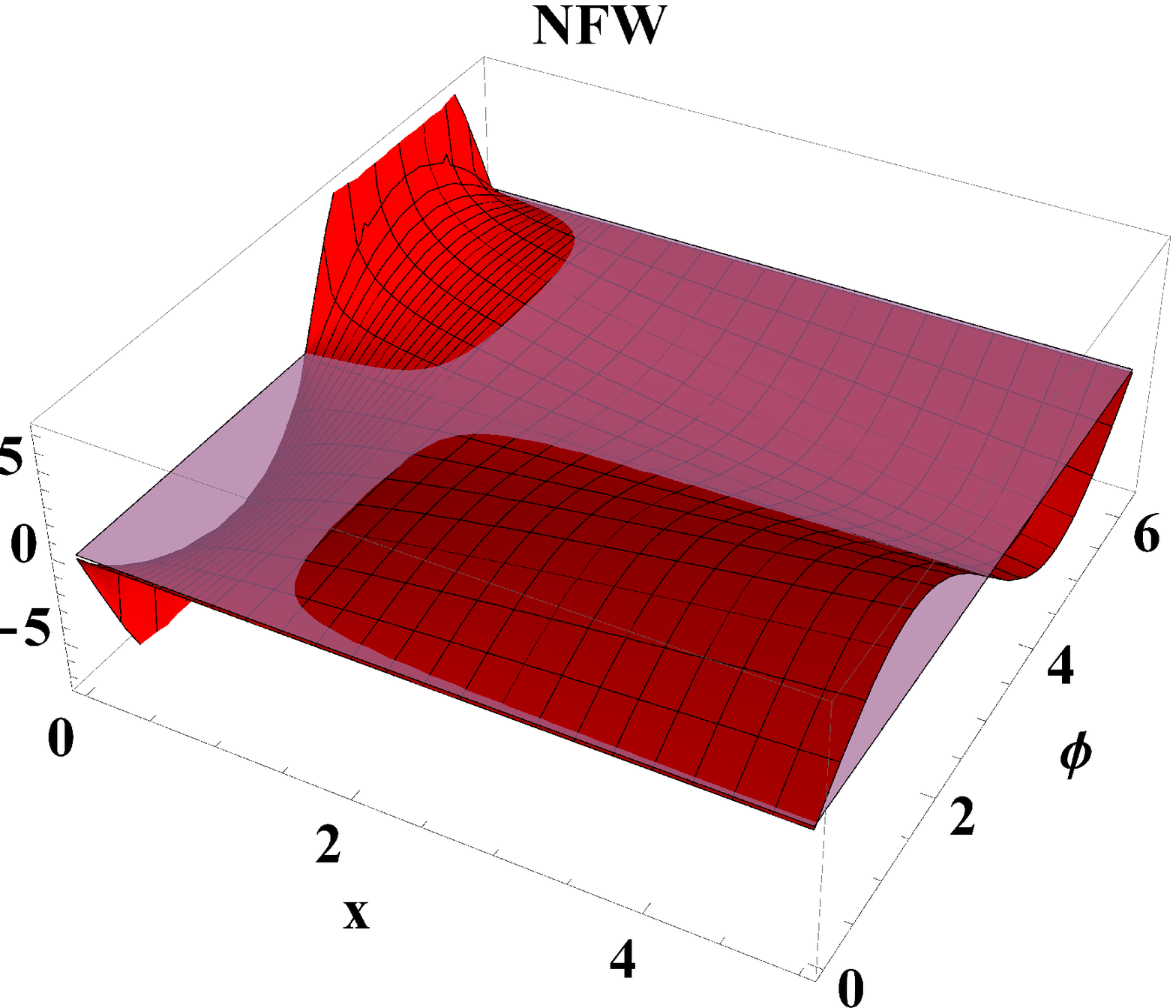} \\[0.5ex]
  \small (c) & \small (d) \\[0.5ex]
\end{tabular}
\caption{Schematic of the surfaces representing the solutions of Eq. (\ref{nsurface}) for each profile on $z_{0}=1$ and $n=0.2$. The pink plane intersecting each surface is the $P=0$ plane. Observe that although the profiles present similar behaviour, the imaging formation process doesn't begin in the same values of $x$ (neither the values of $\phi$ are the same). In fact, for the BEC profile, there are regions (small ``humps'' in the $\pi<\phi<2\pi$ zone) where there are two images corresponding to each angle; however, because of the constraints (\ref{Constraint1}) and (\ref{Constraint2}), those points don't belong to the solutions of Eq. ($\ref{Xsurface}$); for the NFW profile, there are divergences for all the values near to $x=0$.}
\label{profilesn1z1}
\end{figure*}
Hence, the image for each fringe on the plane $z=0$ is
\begin{equation}\label{gravitofringe}
\begin{array}{l}
\displaystyle T_{x}(x)=\pm x\,\,\sqrt{1-\frac{n^{2}}{(x-z_{0}\alpha_{*}(x))^{2}}},\\[3ex]
\displaystyle T_{y}(x)=x \,\,\frac{n}{x-z_{0}\alpha_{*}(x)}.
\end{array}
\end{equation}
For describing the images in the refraction process, it is necessary to find the values of $x$ that satisfy Eqs. (\ref{Xsurface}) and (\ref{nsurface}), to later evaluate those solutions into Eqs.(\ref{gravitofringe}). Because in all cases the deflection angle $\alpha_{*}(x)$ is (in general) not an invertible function, these solutions are obtained numerically.\\
We use as a guide Eq. (\ref{nsurface}) and we interpret it as a family of surfaces labeled by the $n$ index, that are functions of $x$ and $\phi$ (see Figs. \ref{profilesn1z1}). The intersections of these surfaces with the plane $P=0$ are the possible values of $x$ and $\phi$ that satisfy Eq. (\ref{nsurface}), and the plots help us to identify on which positions on the lens plane the corresponding images are formed. However, not all the $x$ values are solutions of Eq.(\ref{Xsurface}), so we must consider that Eqs. (\ref{Xsurface}) and (\ref{nsurface}) are not independent (see the constraints (\ref{Constraint1}) and (\ref{Constraint2})).\\
From the information provided by Eqs (\ref{Xsurface})-(\ref{gravitofringe}) we obtain the relation
\begin{equation}\label{restriction}
  x_{s}-z_{0}\alpha_{*}(x_{s})=(\pm)_{s}\sqrt{{X_{s}^{2}+n^{2}}},
\end{equation}
where $(\pm)_{s}=X_{s}/|X_{s}|$ and $X_{s}$ can be taken for positive or negative values (depending of the position of the analyzed fringe). Observe that the $x_{s}$'s are simply, solutions of the lens equation with the restrictions imposed by $n$ and $X_{s}$.\\
Therefore the $x_{s}$'s solutions are obtained numerically from  Eq. (\ref{restriction}), and these solutions are parameterized by the $X_{s}$ values, for later evaluate them into Eqs.(\ref{gravitofringe}). Using these results we arrive to the equations
\begin{equation}\label{evaluatedfringe}
\begin{array}{l}
\displaystyle T_{x}(x_{s})=\pm \,\,x_{s}\,\,\sqrt{1-\frac{n^{2}}{(x_{s}-z_{0}\alpha_{*}(x_{s}))^{2}}},\\[3ex]
\displaystyle T_{y}(x_{s})=(\pm)_{s}\,\, x_{s} \,\,\frac{n}{x_{s}-z_{0}\alpha_{*}(x_{s})},
\end{array}
\end{equation}
and these are the equations for the image of the fringe parameterized by the $x_{s}$'s solutions, which also encode the information of the $L_{a}\leq X_{s}\leq L_{b}$ interval. That is, Eqs. (\ref{evaluatedfringe}) represent the image of the selected fringe, builded through the comparison of the $x_{s}$ solutions of the lens equation (with the restrictions imposed by $n$ and $X_{s}$ through the constraint \ref{Constraint2}) and the image of such solutions evaluated on the lens equation, that appears in the denominators of $T_{x}$ and $T_{y}$.\\
In this case, because the $\lambda$ parameter encodes the information related with the cosmological distances (the distance along the refracted light rays), we fix the values of $\lambda$ to analyze the evolution of the imaging formation process, because from Eqs. (\ref{strongjacob}), it can be seen that the strong lensing conditions depend directly on this parameter, and the images of the following section are obtained in this manner.

\section{Imaging formation for DM profiles}
Using the functions given in table \ref{TAB} and Eqs. (\ref{evaluatedfringe}), the images generated with each deflection angle are computed.\\
The relevant data is encoded in the $\lambda$ parameter \cite{Herrera:strong}, and there are two options to study the behavior of the images through variations of such quantity. One, is to input values for the distance factor $d_{s}/(d_{l}d_{ls})$ by using specific values obtained from cosmological data (for example from SLACS, LSD and SL2S \cite{Herrera:strong}); the other is to fix the values of $\lambda$ to study the evolution of the rays and caustics, and in turn for studying the images generated in each case. This latter approach is useful for the calculations in our work, because allow us to establish regions for comparing the scale of the Einstein rings and the images there directly from the plots.\\
\subsection{Examples}
Some positions of the source plane and fringes that seems to be interesting for analysis are shown.\\
Although both the $z_{0}$ and the $n$ parameters are continuous, we choose discrete values for this analysis. In this case for $z_{0}=\lambda=1$ and later for $\lambda=2$ to analyze specifically the MSFDM profile.\\
For the fringes, we choose one near the observation axis ($n=0.2$), that represents a linear source defined by the interval $0.3 \leq X_{s} \leq 0.9$, and another placed farther from the observation axis (for $n=0.8$) on the interval $-0.5<X_{s}<0.5$ (both on the source plane, see Figs. \ref{Fringe26} and \ref{Fringe81} ), to analyze what are the differences between these cases.\\

\subsubsection{Rays and caustics}
\begin{figure}[h]
  \centering
  \begin{tabular}{c}
  \includegraphics[width=60mm]{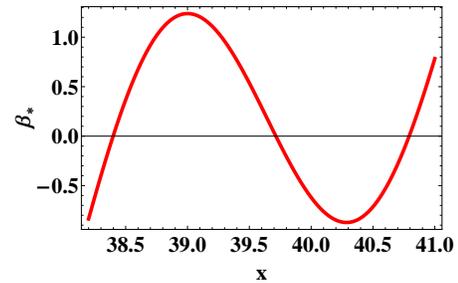}\\[0.5ex]
  \small (a)\\[2ex]
  \includegraphics[width=60mm]{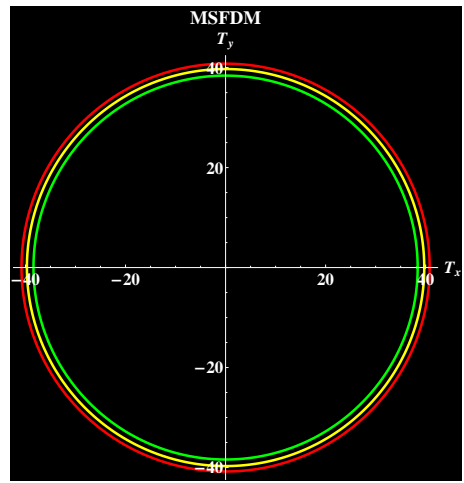}\\[0.5ex]
  \small (b)
  \end{tabular}
  \caption{(a)Close up for the roots of the lens mapping $\beta_{*}$ with $\lambda=2$ for the MSFDM profile. Observe that these roots represent three concentric Einstein rings. (b) Plots on the lens plane of the Einstein rings associated with such roots; observationally these rings could be interpreted as an unique thick ring for that position of the source plane.}\label{CloseTriple}
\end{figure}

\begin{figure*}[h]
\begin{tabular}{c c}
  % after \\: \hline or \cline{col1-col2} \cline{col3-col4} ...
  \includegraphics[width=62mm]{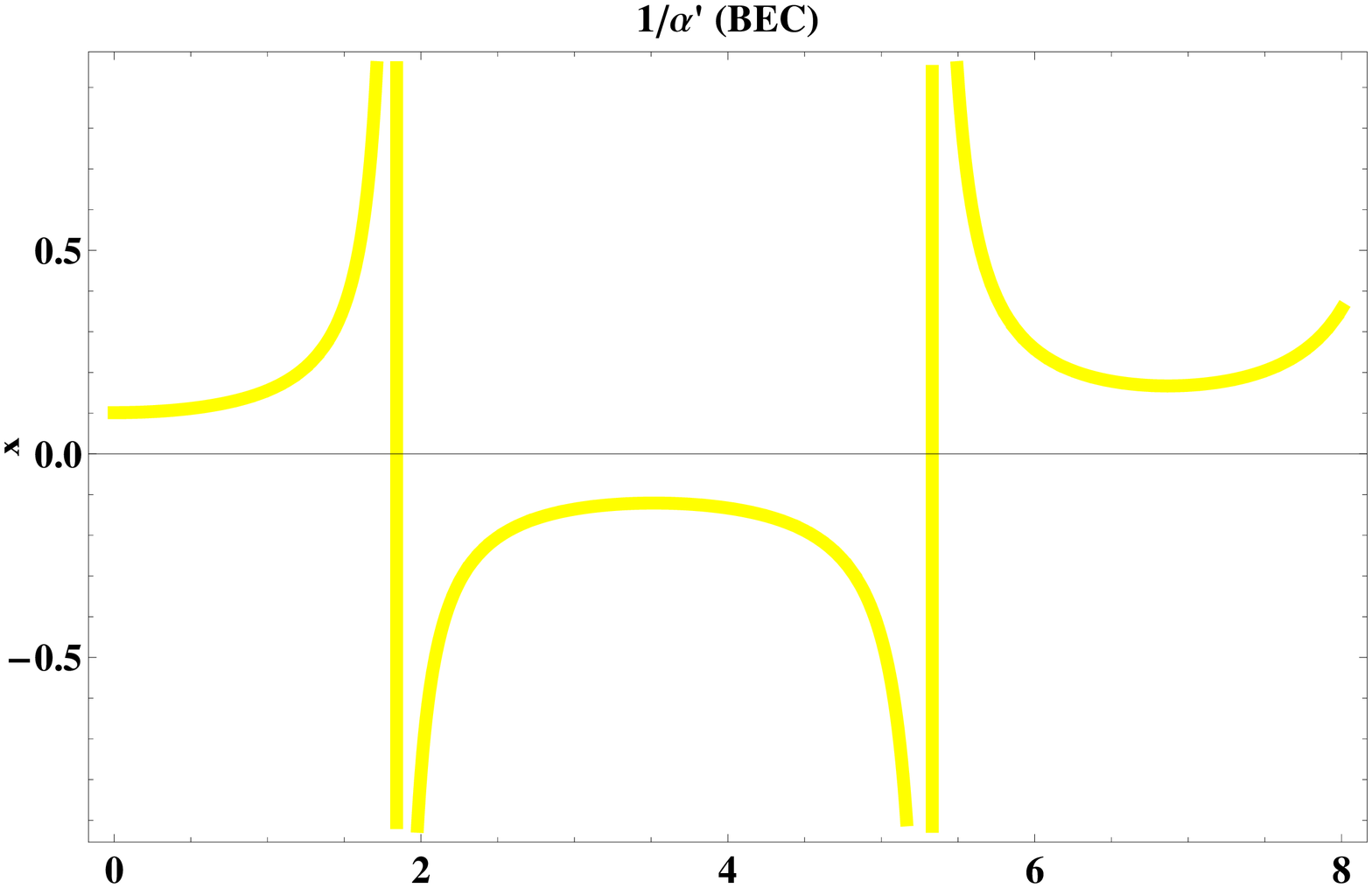} & \includegraphics[width=64mm]{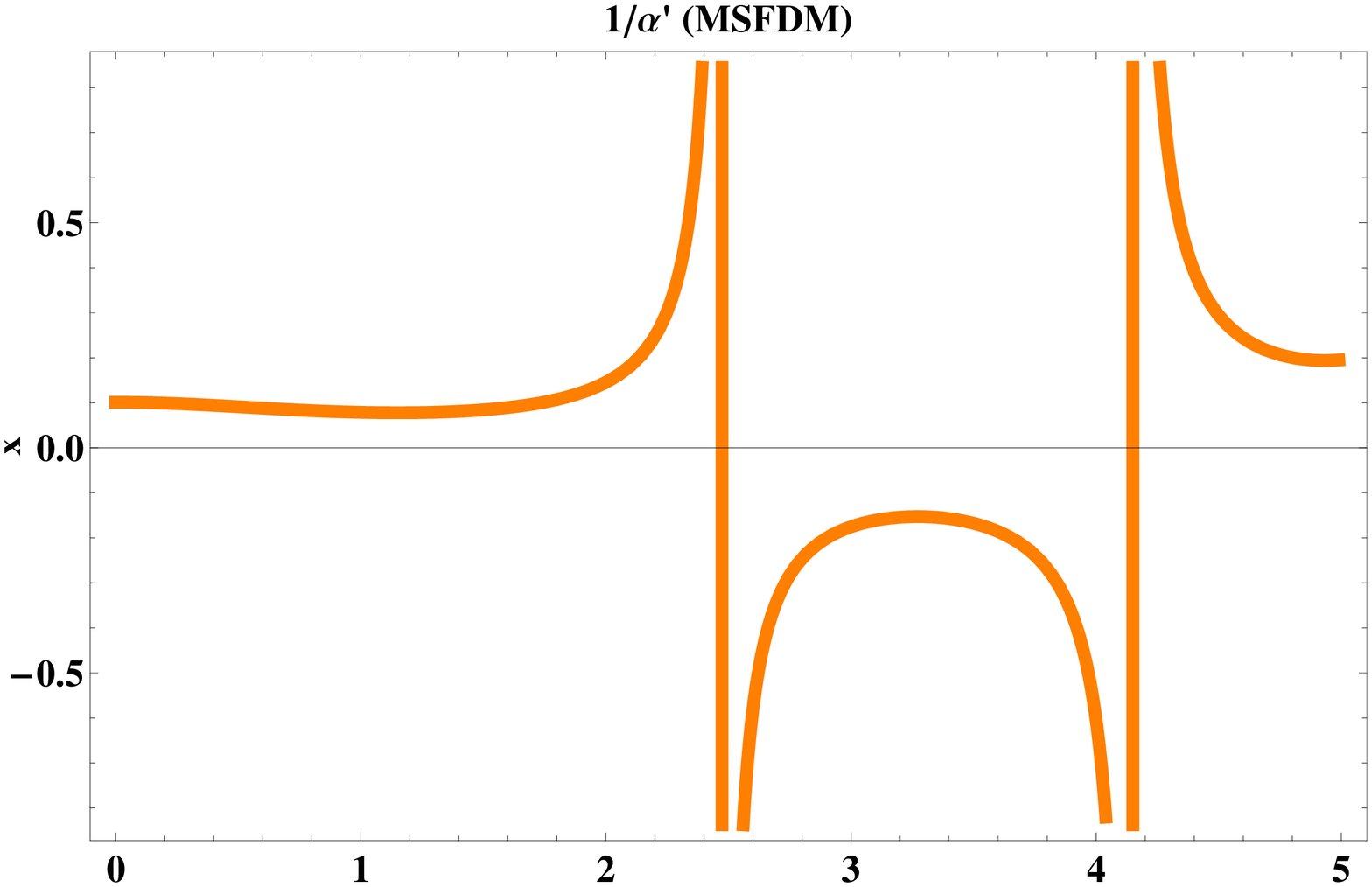} \\[0.5ex]
  \small (a) $0\leq x<1.841$ & \small (b) $0\leq x<2.473$ \\[0.5ex]
  \includegraphics[width=68mm]{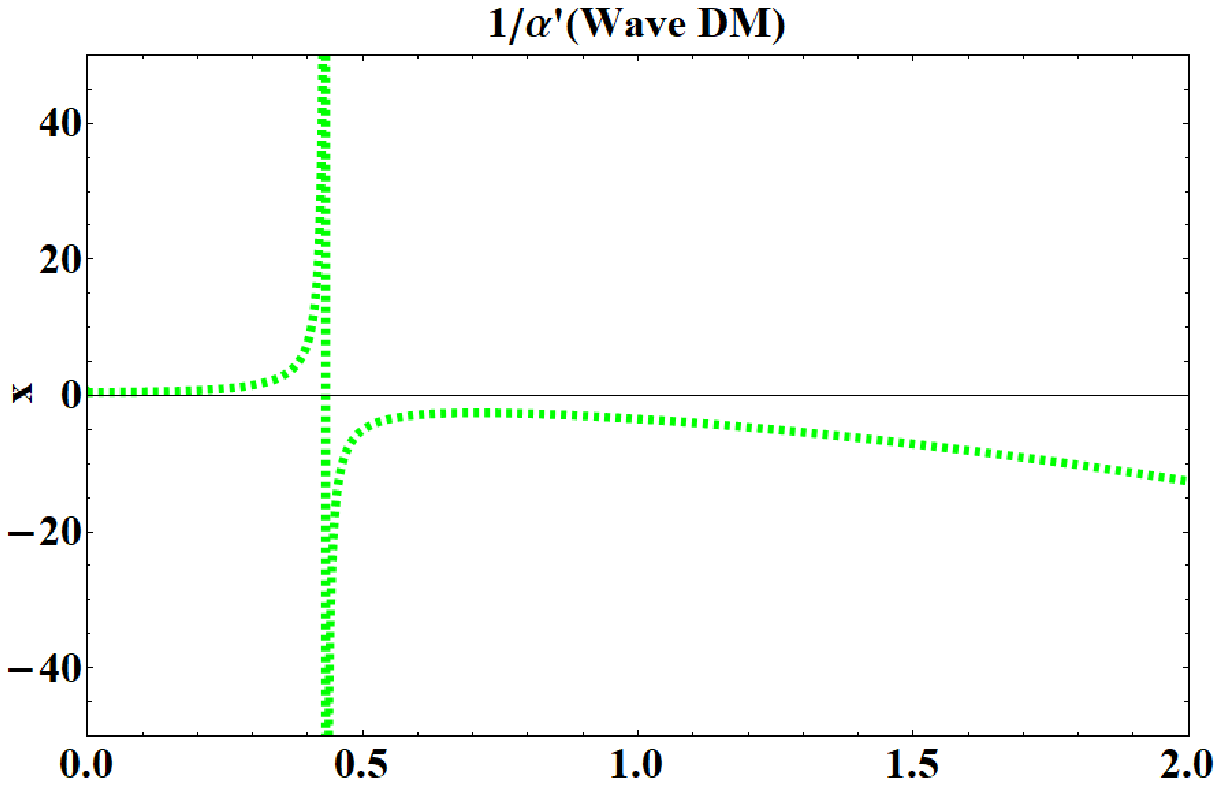} & \includegraphics[width=70mm]{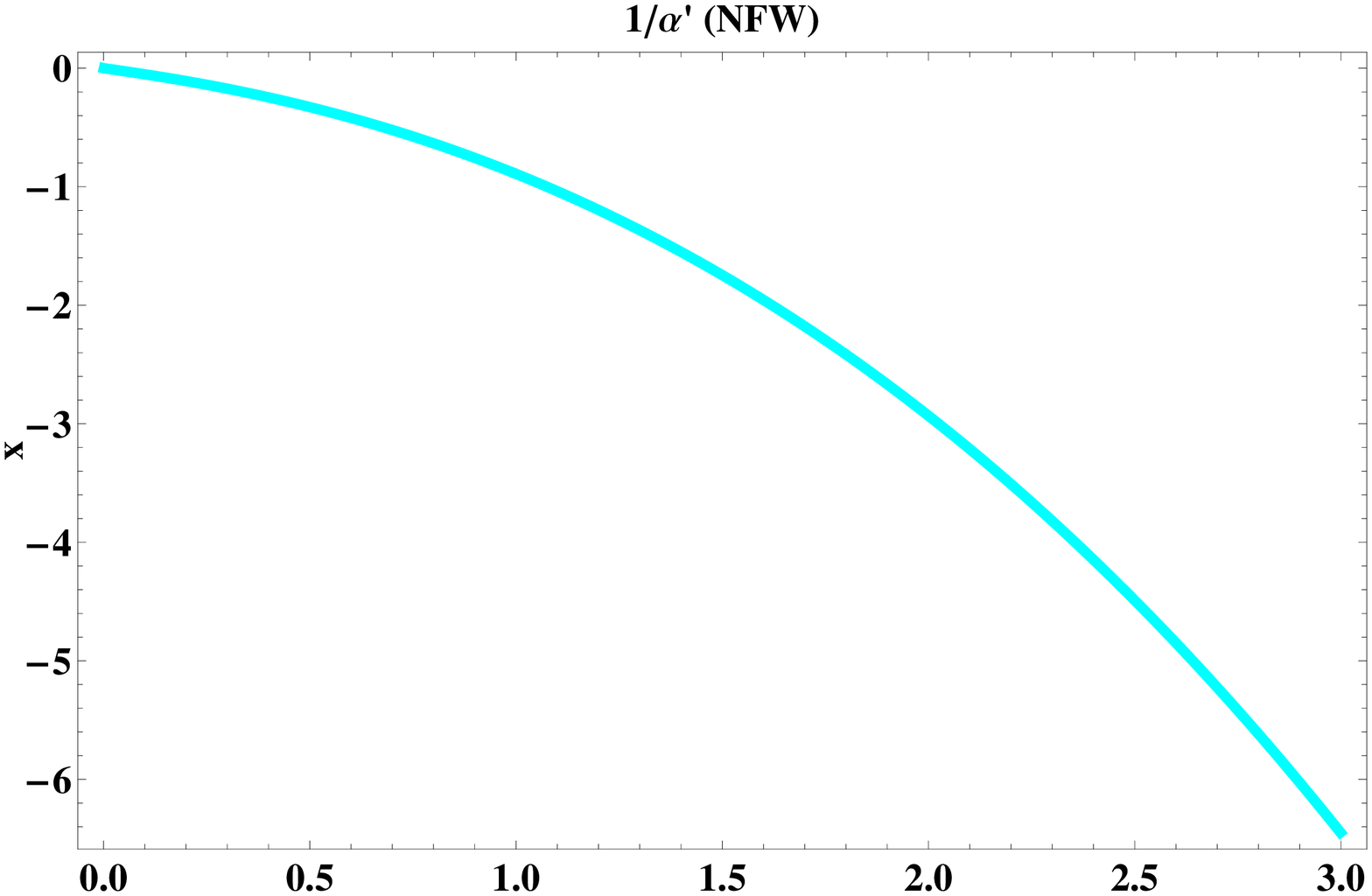} \\[0.5ex]
  \small (c) $0\leq x < 0.432$  & \small (d) $/$ \\[0.5ex]
\end{tabular}
\caption{Plots representing the function $\mathbf{X}_{-}\cdot{\hat{\bm{z}}}$ from Eqs. (\ref{causticBr}), which gives information for the regions where the $z$ component of the $\mathbf{X}_{-}$ branch is real. Below each plot, it is noted the interval where the plot is positive, \textsl{i.e.}, the region where the caustics are real; for the BEC and MSFDM profiles, only the first interval where the lenses present strong lensing is noted, however there are an infinitude of regions where there is strong lensing.\\
An important observation is that for the NFW profile, the revolution surface that represents the $\mathbf{X}_{-}$ branch of the caustic in that case, is virtual.}
\label{virtualzone}
\end{figure*}

\begin{table}[h]

%\caption*{}\\
{\begin{tabular}{| l | c | c | }\hline
\multicolumn{3}{|c|}{\textbf{Central caustic regions}}\\\hline
\textsl{Profile}    & $X_{i+}$ & $X_{f+}$\\[1ex]\hline
%\textbf{Burkert}    &    $2/\pi^{2}\approx 0.203$        & $0.219$  \\[1ex]
\textbf{BEC}        &    $1/\pi^{2}\approx 0.101$        & $\infty$ \\[1ex]
\textbf{MSFDM}      &    $1/\pi^{2}\approx 0.101$        & $\infty$ \\[1ex]
%\textbf{PI}         &    $1/\pi^{2}\approx 0.101$        & $\infty$ \\[1ex]
%\textbf{Spano}      &    $1/2\pi\approx 0.159$           & $\infty$ \\[1ex]
\textbf{Wave DM}    &    $2^{7}7!/(13!!\pi^{2})\approx 0.484$   & $\infty$ \\[1ex]
\textbf{NFW}        &    $0$                             & $\infty$
\\\hline
\end{tabular}}
  \caption{Beginning ($X_{i+}$) and end ($X_{f+}$) of the central branch of the caustic for each profile. Using Eqs. (\ref{causticBr}) it is obtained the nearest value of this caustic along the observation ($z$) axis. Observe that for all cases, $X_{i+}=\lambda_{cr}$ which means that the caustic evolves as the $\lambda$ parameter does.}\label{centralXbranch}
\end{table}
In the region between the source and the lens plane, we can develop the ray tracing for each of the profiles using Eq. (\ref{Xreal}).\\
For the caustic branches produced by each lens, we use Eqs. (\ref{causticBr}) to obtain the projection of such branches along the $z$ axis. But first we must remember that not all the regions of the caustic are real, and this condition arises from the $z$ component on Eqs. (\ref{causticBr}). That is, it is important to find where the function that defines this component is positive; such functions in each case are difficult to solve, so we plot those function to directly search for the values where the caustic is real (see plots \ref{virtualzone} ).\\
From the plots representing the rays and caustics generated for each profile, there are differences belonging to each caustic, that have direct consequences on the formation of Einstein rings.\\
For the BEC profile not all the segments of the caustic are real, due to the alternating positivity and negativity of the Bessel function $J_{0}(x)$; however, the positive regions do contribute to the caustic and the central branch evolve up to infinity and the negative regions contribute to a virtual branch. The caustic is of the cusp type \cite{Petters:singtheory}.\\
The MSFDM profile, presents a (small) caustic of the butterfly type, but in this case the central branch extends to infinity. Hence, here also are present Einstein rings but in this case there is the possibility that the MSFDM profile produce multiple Einstein rings for a fixed $\lambda$. For example, for $\lambda=2$ there are three concentric Einstein rings with radii $x_{1}\approx 38.4$ (green), $x_{2}\approx 39.71$ (yellow) and $x_{3}\approx 40.79$ (red), respectively (see Fig. \ref{CloseTriple}). Because of the closeness of these rings, observationally this configuration could be interpreted as an unique thick ring.\\
For the Wave DM profile, the caustic is also of the cusp type, but the rays converge slowly, in comparison with  the previous cases. It can be seen  that the shape of the caustic is a cone, but the domain of the revolution surface is finite (see plots \ref{virtualzone}). \\
Finally, for the NFW profile, the revolution surface of the caustic is always virtual, and the only real part is the central branch. Because the domain of this branch is all the real positives (NFW halos can induce strong lensing for any $\lambda$ \cite{Burkert:glburhalos}), this implies that this profile always produces Einstein rings due to the fact that this branch always is in contact with the source plane.\\
That is, an important conclusion provided by the previous information is that the central branch of the caustic gives rise to Einstein rings: if the source plane is placed in a region where the central branch touches such plane, Einstein rings appear, that are solutions of Eq. (\ref{EinRingBeta}).

\newpage
\LTcapwidth=\textwidth
\begin{longtable*}[h]{|c|c|c|}\hline
\multicolumn{3}{|c|}{a) \textbf{BEC}}\\\hline
\includegraphics[width=50mm]{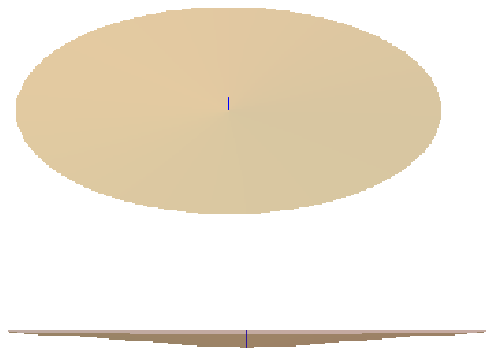} & \includegraphics[width=22mm]{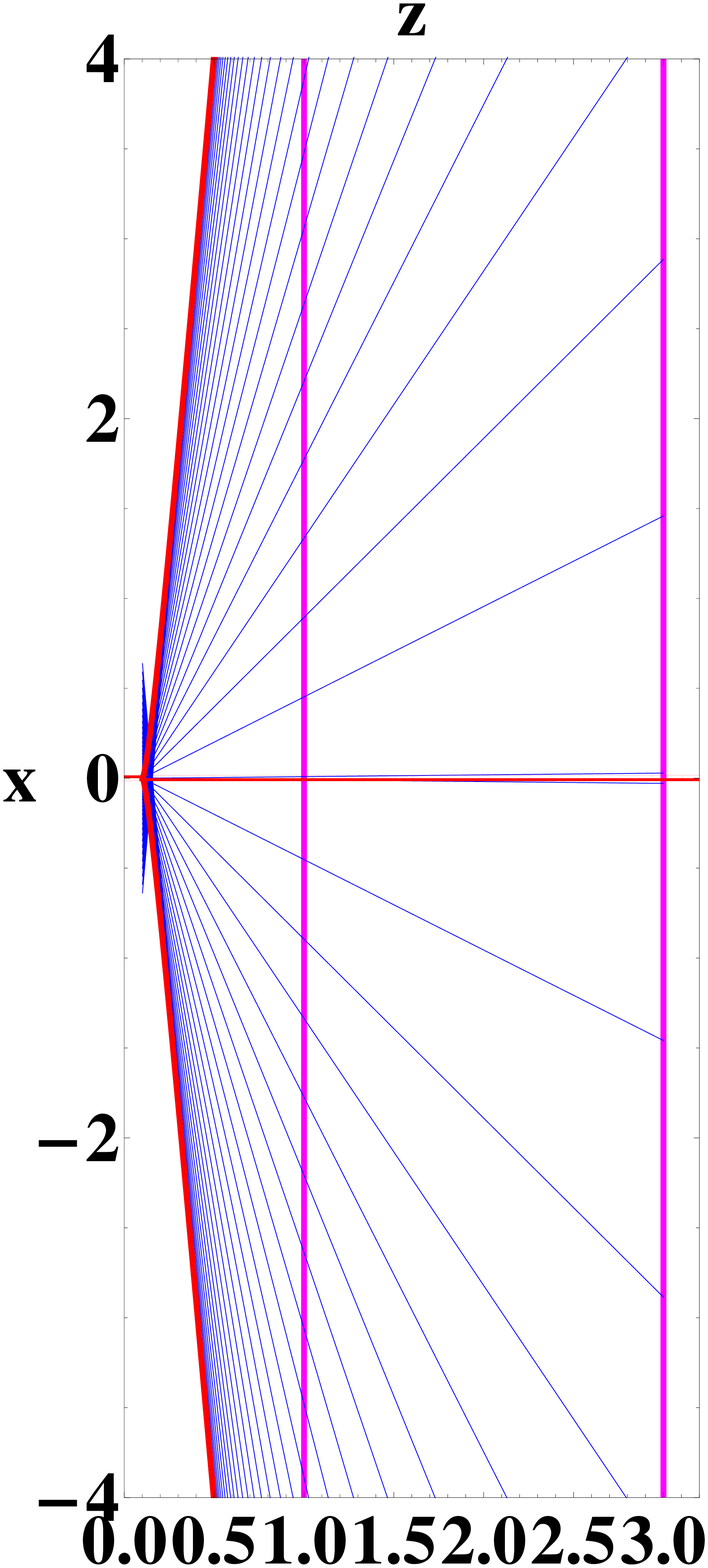} & \includegraphics[width=34mm]{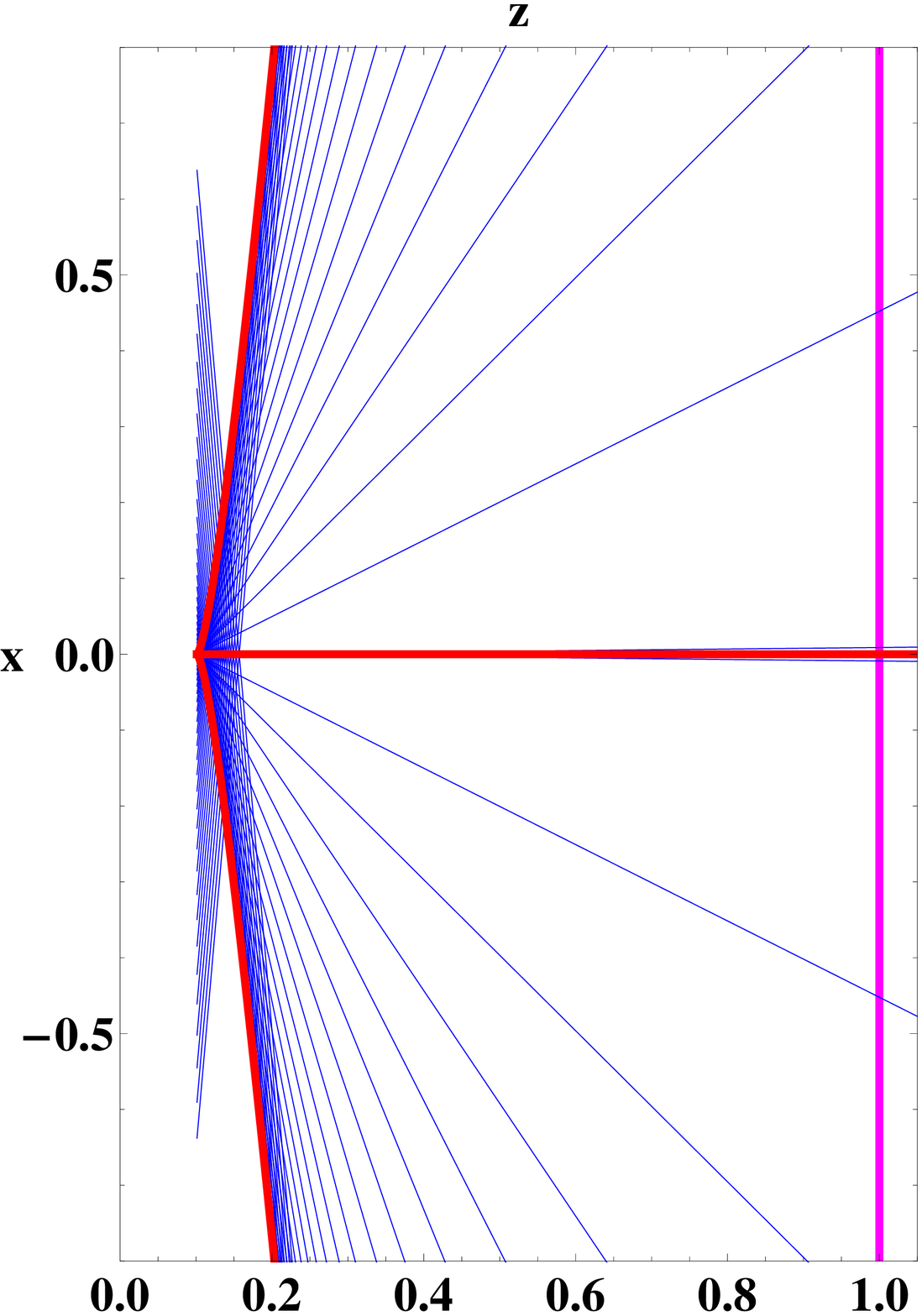} \\[3ex]\hline
\multicolumn{3}{|c|}{b) \textbf{MSFDM}}\\\hline
\includegraphics[width=45mm]{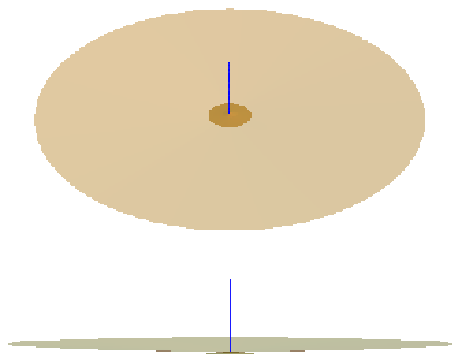} &\includegraphics[width=25mm]{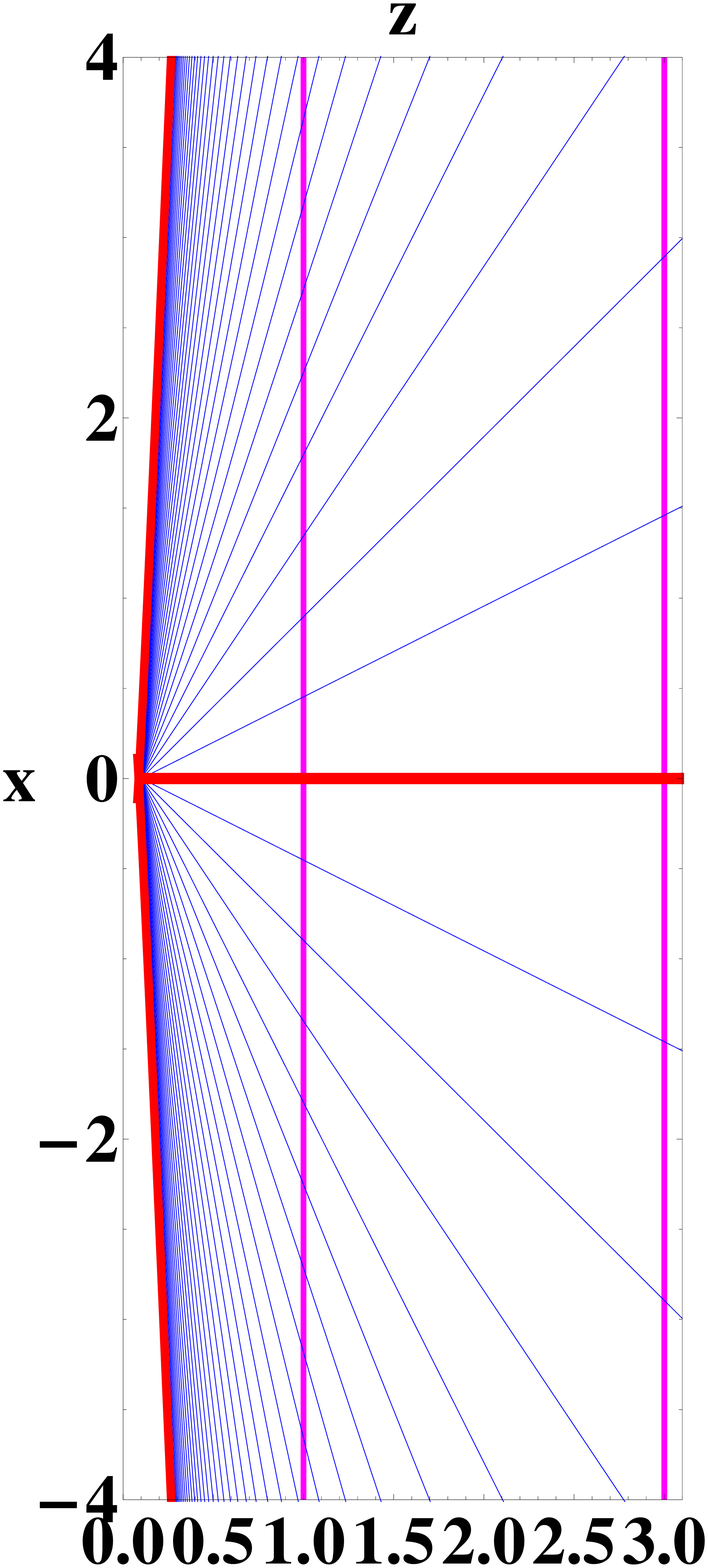} & \includegraphics[width=25mm]{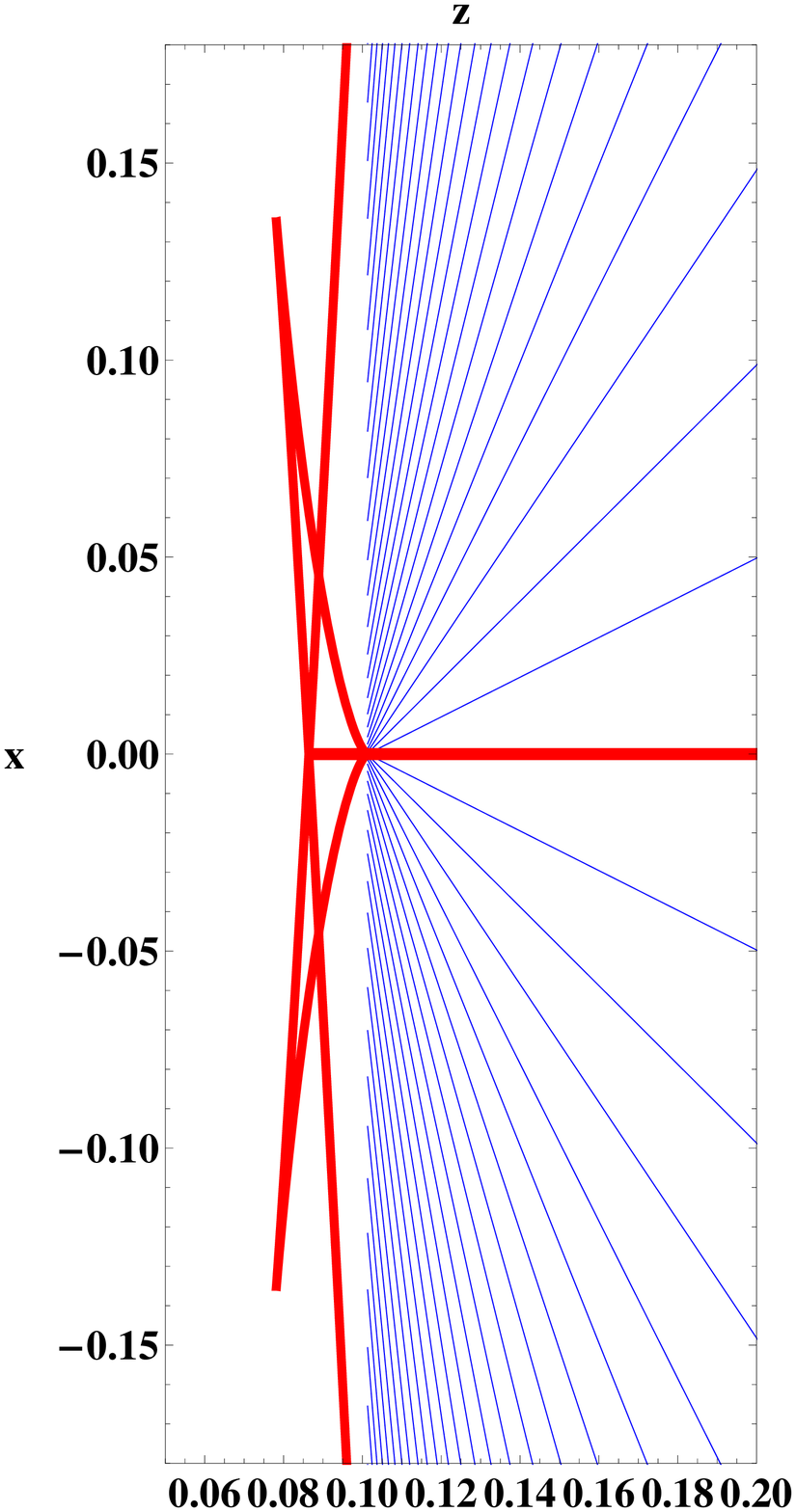} \\[3ex]\hline
\multicolumn{3}{|c|}{c) \textbf{Wave DM}}\\\hline
\includegraphics[width=35mm]{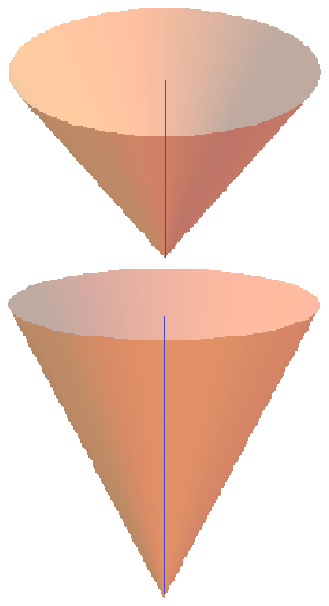} & \includegraphics[width=40mm]{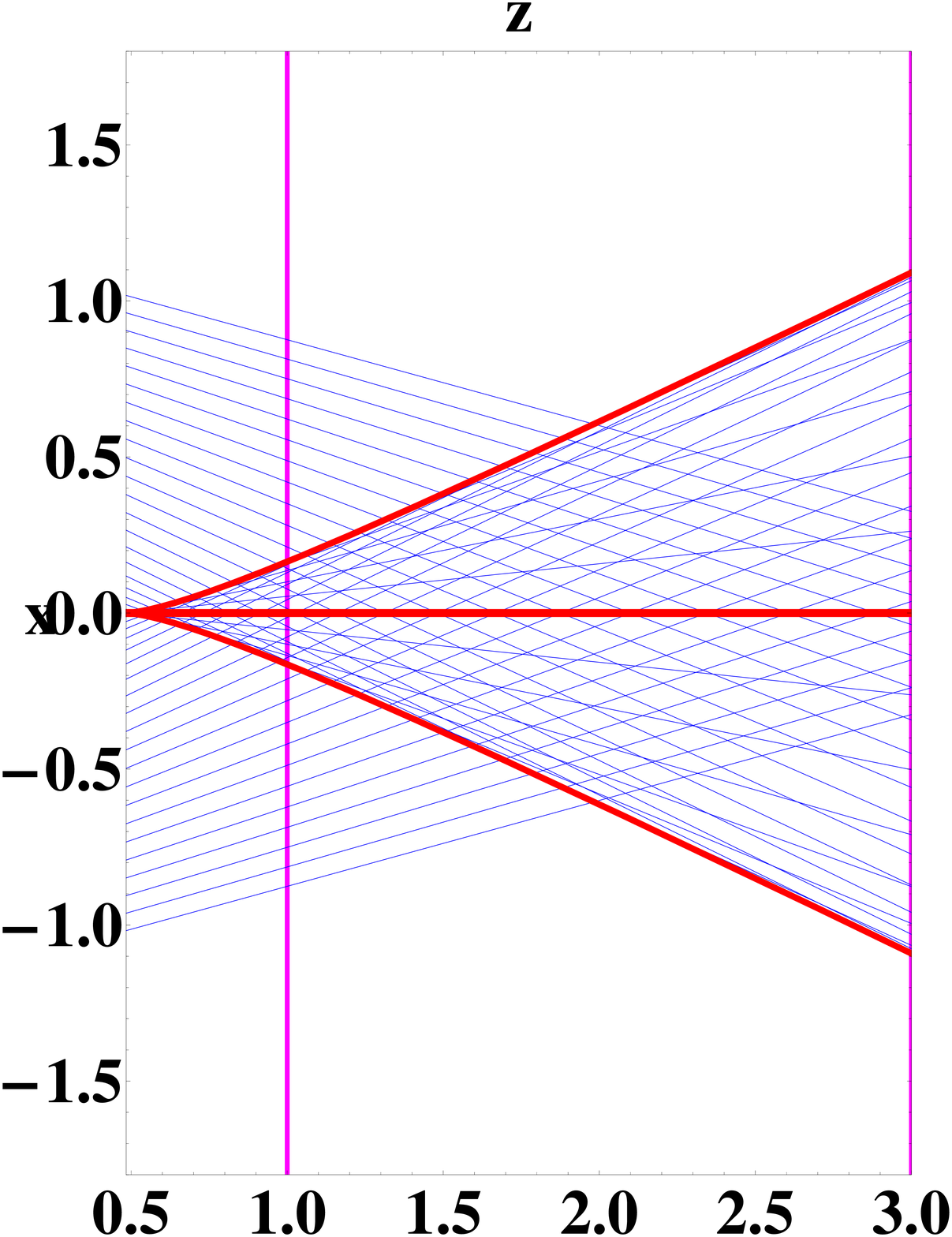} & \includegraphics[width=42mm]{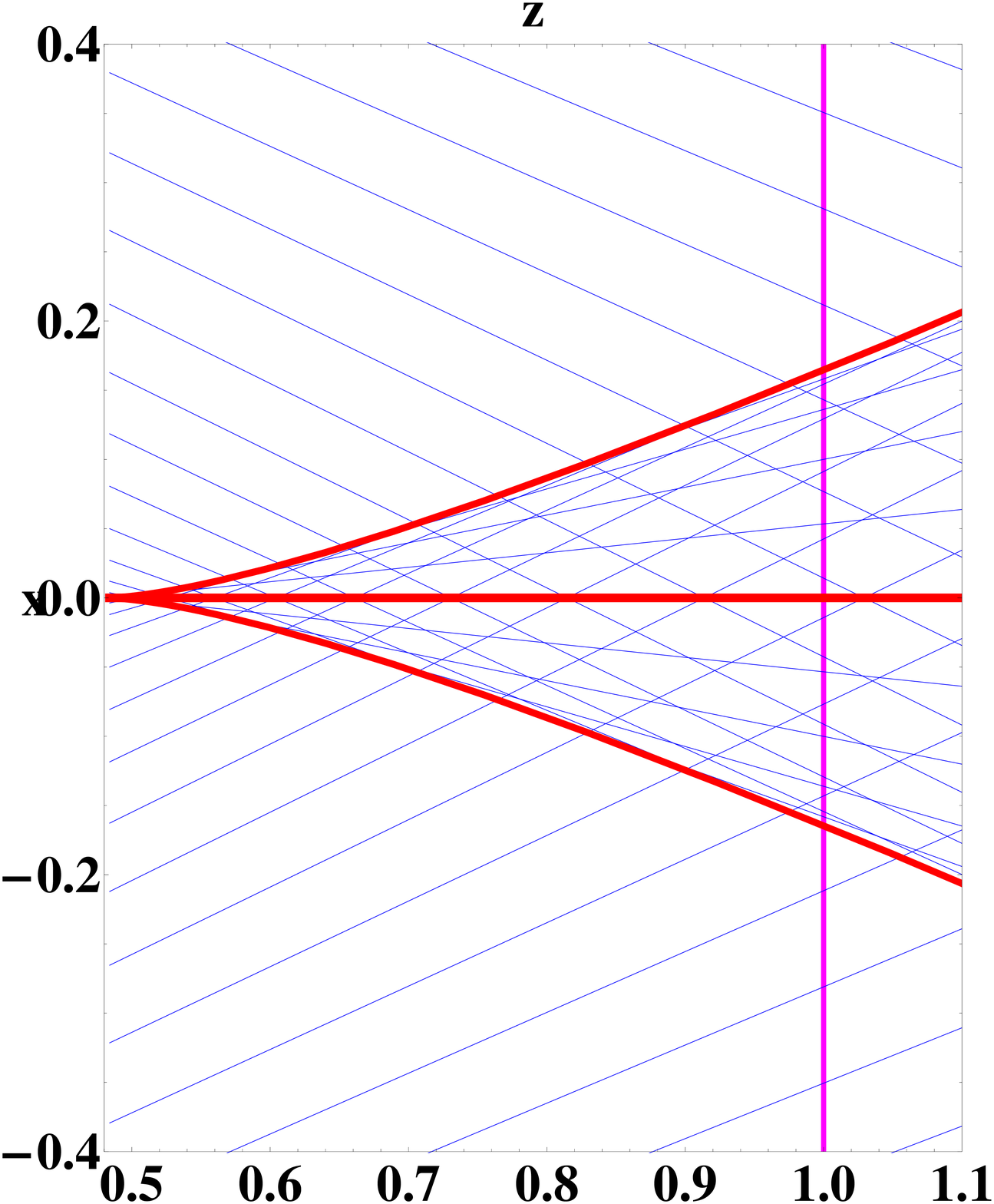} \\[3ex]\hline
\multicolumn{3}{|c|}{d) \textbf{NFW}}\\\hline
\includegraphics[width=35mm]{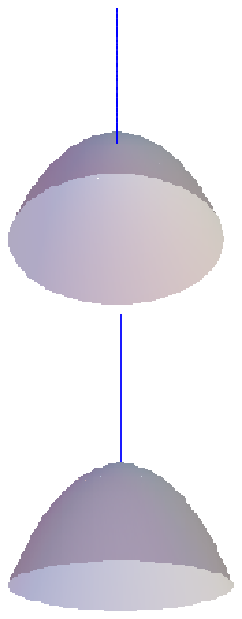} & \includegraphics[width=22mm]{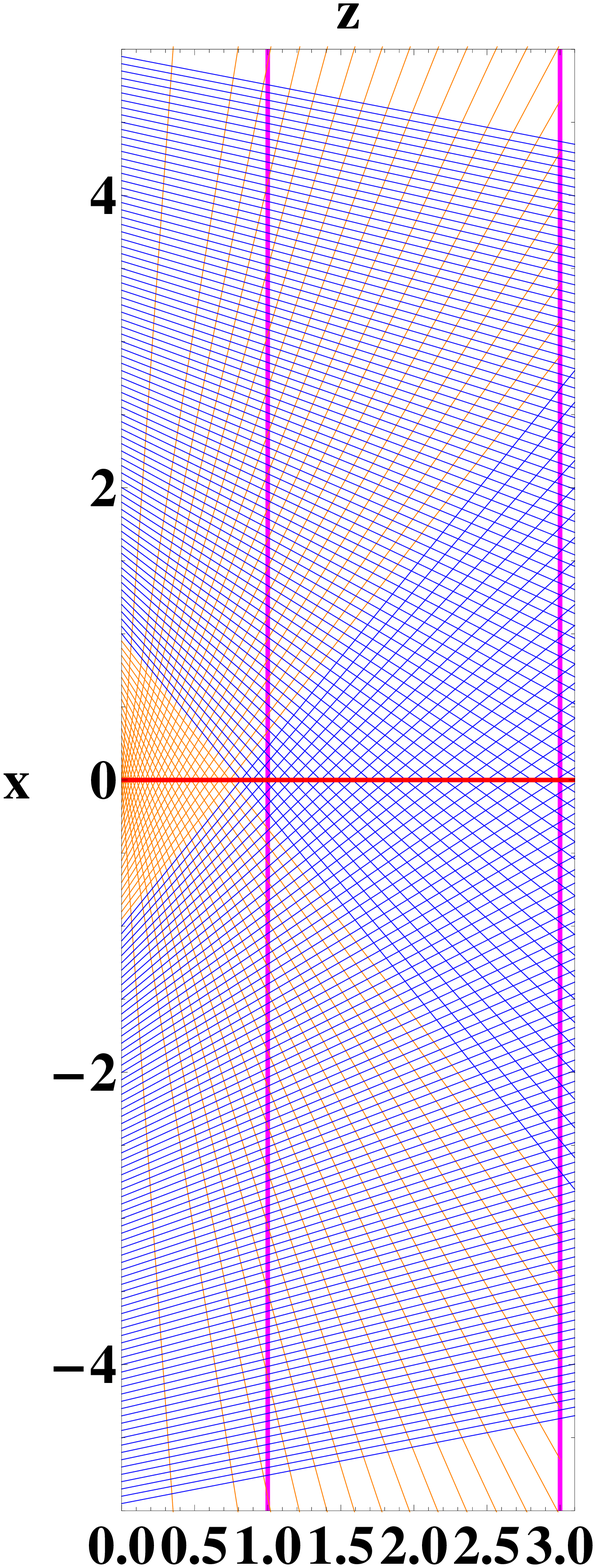} & \includegraphics[width=28mm]{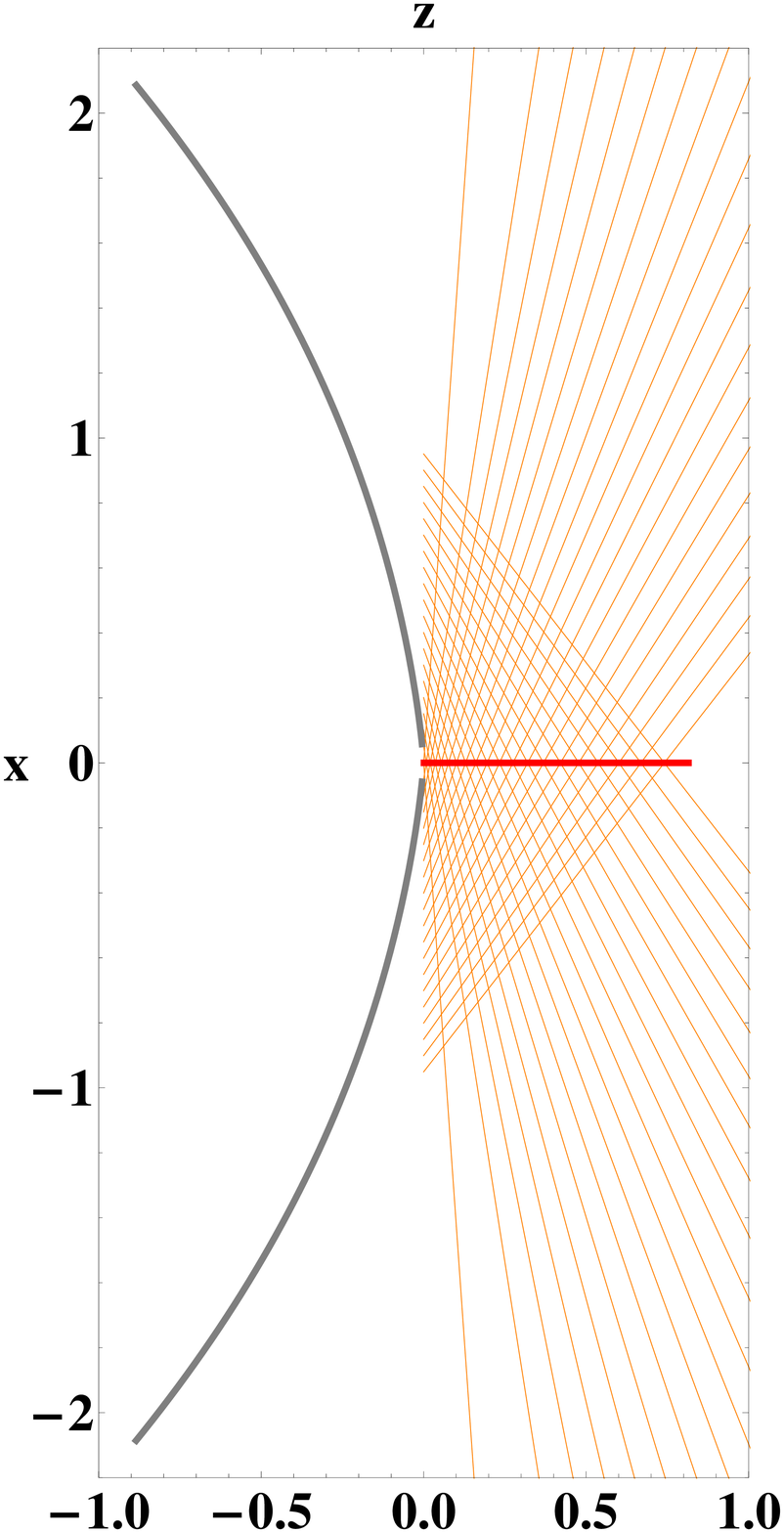} \\[3ex]\hline
\caption{Schematics of the caustics generated for each profile. First column: surfaces that represent the caustic (top: a 3D view; bottom: a frontal view); the orange surface is the $\mathbf{X}_{-}$ branch which is a surface of revolution around the observation axis (except for NFW which is represented by a gray surface), meanwhile the blue central line is the $\mathbf{X}_{+}$ branch, that is always a segment of line along the $z$ axis. Second column: rays generated by each deflection angle $\alpha(x)_{*}$ (in blue), a section of the caustic on the $xz$ plane (red) and the source planes for $\lambda=1$ and $\lambda=3$ (purple lines). Third column: close up for the caustic regions where each profile presents a different behaviour; observe that BEC and Wave DM have caustics of the cusp type, meanwhile the Multistate profile has a caustic of the butterfly type. An important observation for the NFW profile, is that the only real branch of the caustic is the central branch; the virtual part of the caustic (gray surface and gray curve, respectively) evolves on the negative direction of the $z$ axis, towards the observer.}\label{3Dplots}
\end{longtable*}

\subsubsection{Imaging formation}
\begin{figure*}
\begin{tabular}{c c c}
\includegraphics[width=45mm]{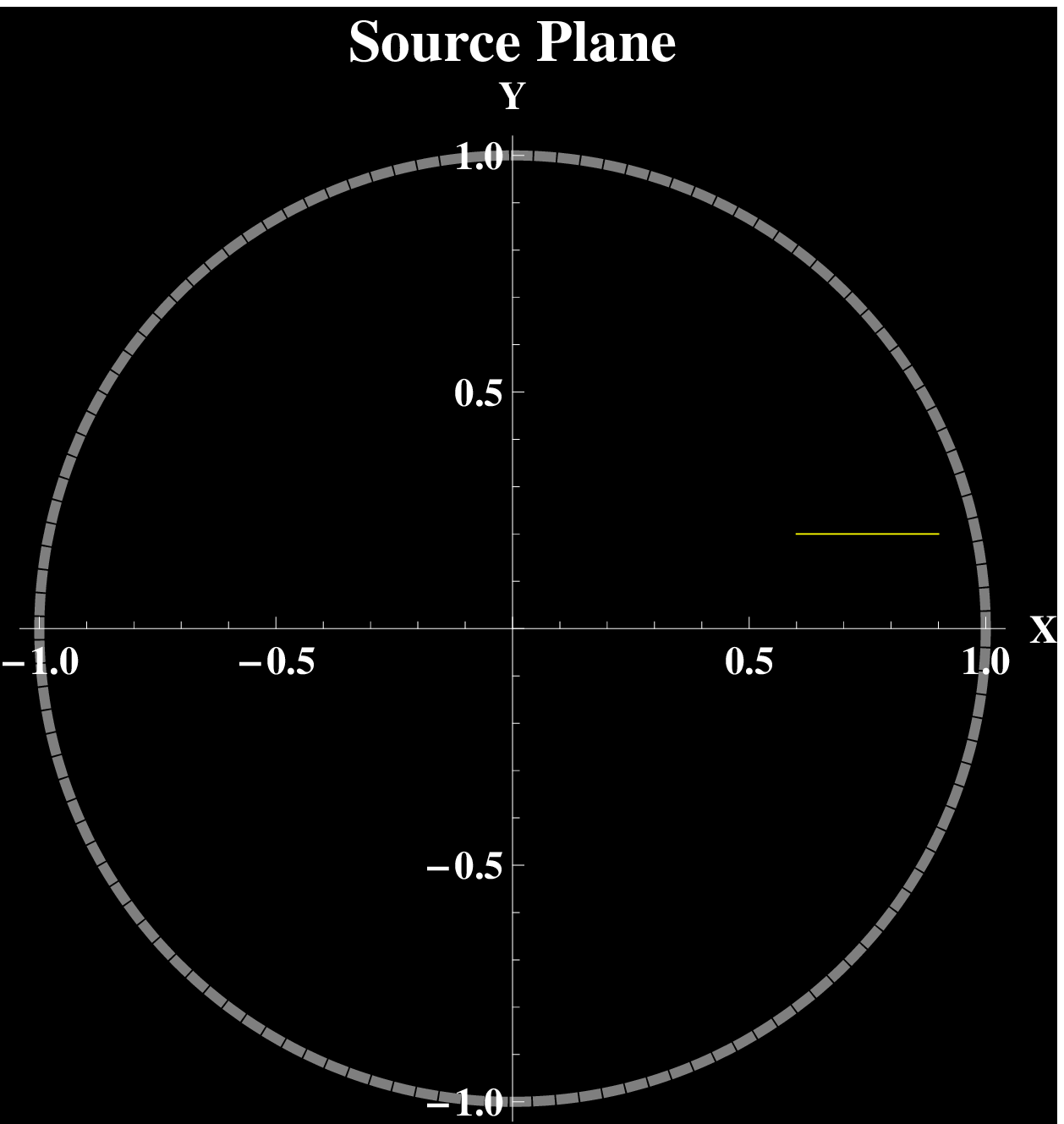} & {} & {}\\[0.5ex]
  \small (1) & {} & {}\\[3ex]
\includegraphics[width=75mm]{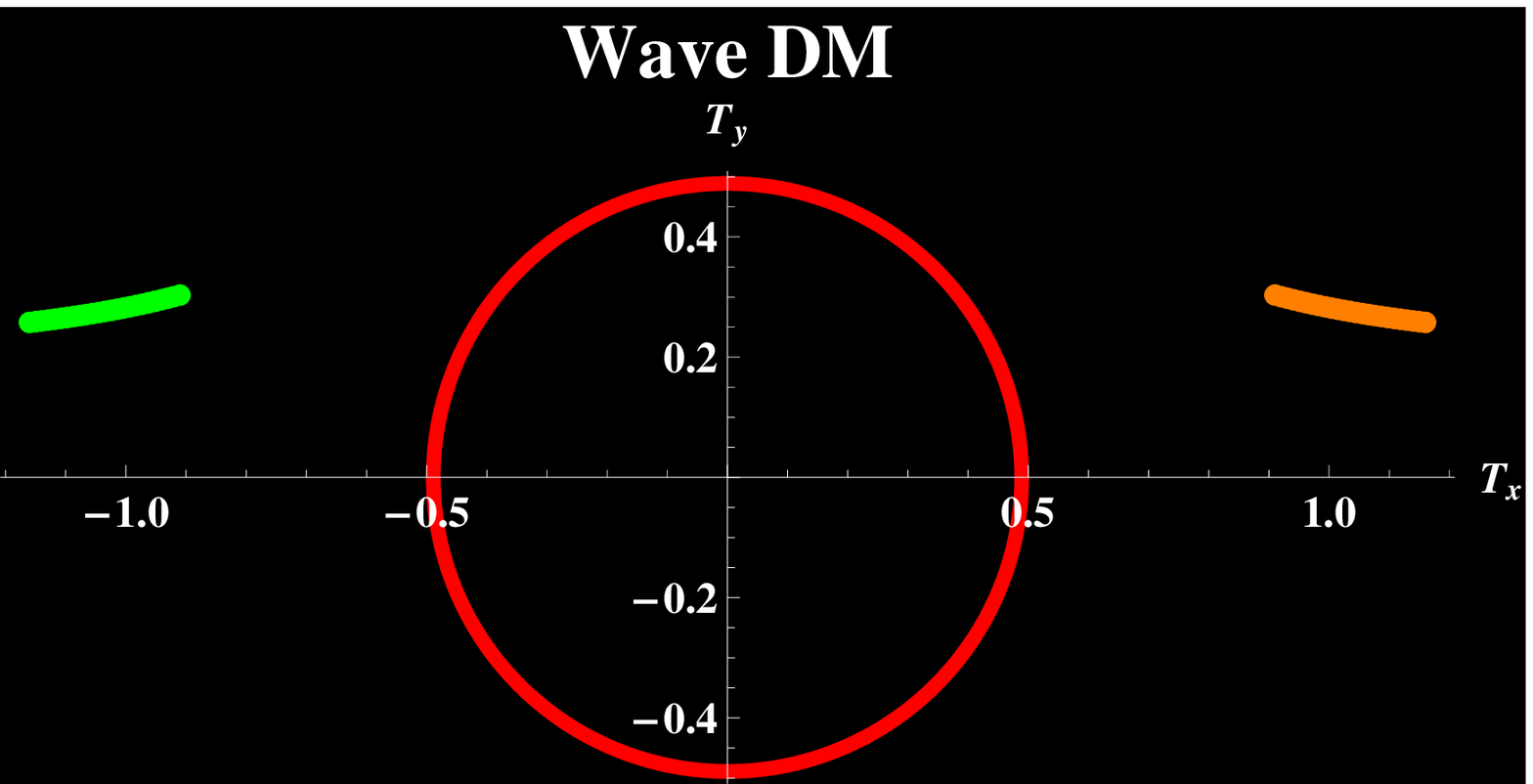} & {} & \includegraphics[width=54mm]{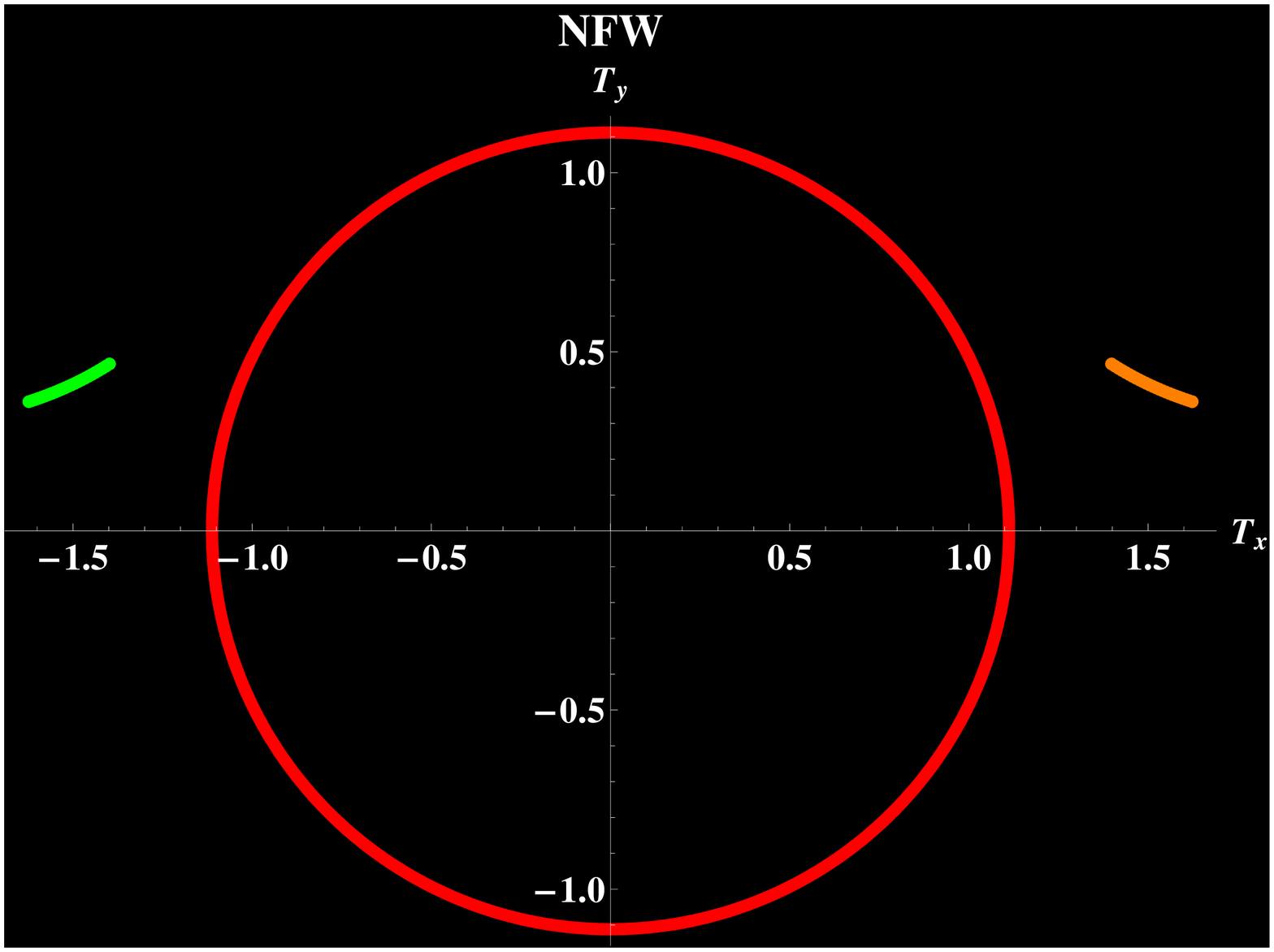}\\[0.5ex]
  \small (a) & {} & \small (b)\\[3ex]
\end{tabular}
\caption{1) A (yellow) fringe that represents a linear source placed at $n=0.2$ with $0.3\leq X_{s}\leq 0.9$, which is mapped with the corresponding deflection angle $\alpha_{*}(x)$ from table (\ref{TAB}I) and using Eqs. (\ref{EinRingBeta}) and (\ref{evaluatedfringe}); the gray circle (of radius 1) is placed to give a scale and orientation for the source. a)-b) Images on the lens plane for some of the studied profiles: Wave DM and NFW. The red circle is the Einstein ring that arise from the contact of the central branch of the corresponding caustic with the source plane at $\lambda=1$.}\label{Fringe26}
\end{figure*}

\begin{figure*}
\begin{tabular}{c c c}
\includegraphics[width=55mm]{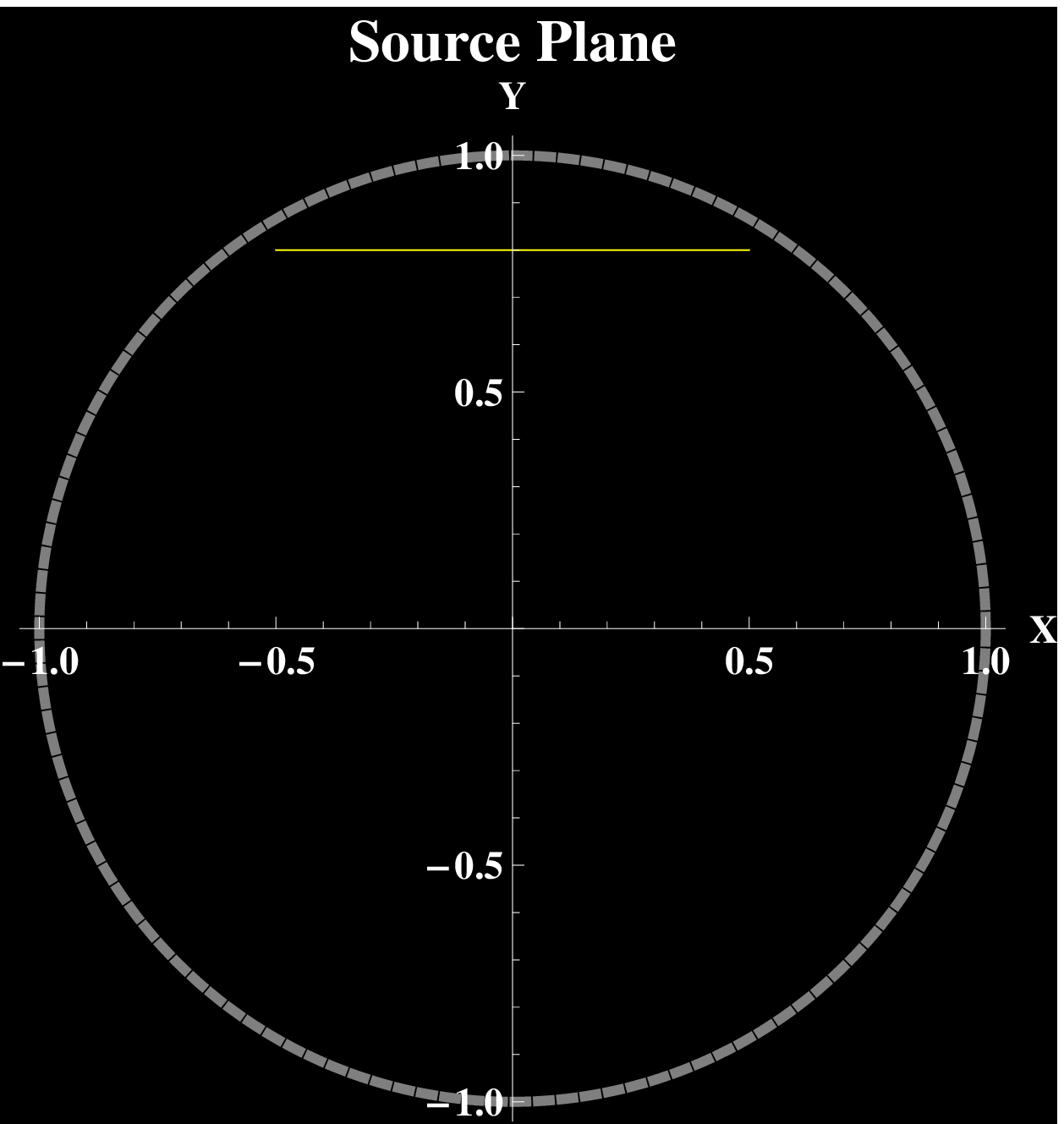} & {} & \includegraphics[width=35mm]{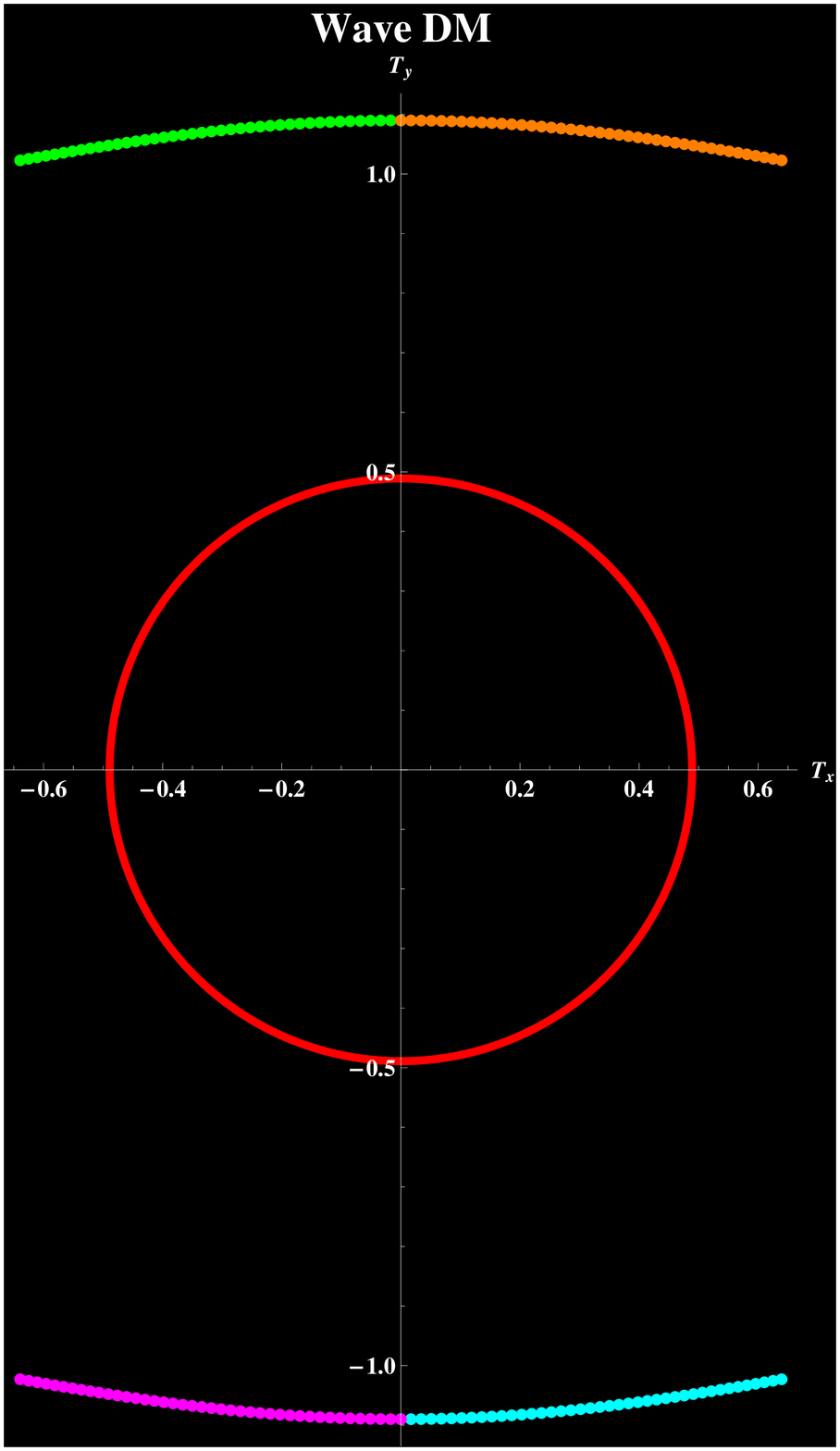}\\[3ex]
\end{tabular}
\caption{A (yellow) fringe representing a linear source placed symmetrically respect to the $T_{y}$ axis at $n=0.8$. Because of the position of the source, in this case Eqs. (\ref{evaluatedfringe}) produce a symmetric image below the $T_{x}$ axis. %For the example given here (Wave DM), we can see that each part of the arclets corresponds to.
}\label{Fringe81}
\end{figure*}

With the caustic regions established, the behaviour of the images mapped by Eq. (\ref{Xreal})is analyzed. Remembering that the $\mathbf{X}$ vector field is in this case, the three dimensional representation of the lens mapping, and the images generated by such mapping are described by Eqs. (\ref{evaluatedfringe}).\\
The strong lensing regions are determined by fixing the values of $\lambda$, for later  using Eq. (\ref{strongjacob}) (see table \ref{strongLENS}). Although the $\lambda$ parameter is a continuous function, in these examples we choose some values of $\lambda$, to analyze the imaging formation for each profile, and for specific sizes and positions of the fringes placed on the source plane.\\
We approach first a source placed very near the observation axis which doesn't cross the vertical axis $T_{y}=0$, for $n=0.2$ and length between $0.6\leq X_{s}\leq 0.9$, and later, another fringe placed symmetrically respect to the $T_{y}$ axis, but farther above from the $T_{x}$ axis for $n=0.8$ with length between $-0.5\leq X_{s}\leq 0.5$ (see Figs. \ref{Fringe26} and \ref{Fringe81} ). Finally, an special case for the MSFDM profile in $\lambda=2$ is shown, to observe the multiplicity of Einstein rings for that position of the source plane. \\
\begin{table}[h]
{\begin{tabular}{| l | c | }\hline
\multicolumn{2}{|c|}{\textbf{Strong lensing regions}}\\\hline
\textsl{Profile}    & $\lambda=1$ \\[1ex]\hline
%\textbf{Burkert}    &    /        \\[1ex]
\textbf{BEC}        &  $1.718<x<3.415$\\[1ex]
\textbf{MSFDM}      & $2.404<x<4.242$\\[1ex]
\textbf{Wave DM}    & $0.245<x<0.489$ \\[1ex]
\textbf{NFW}        &   $0<x<1.112$
\\\hline
\end{tabular}}
  \caption{Strong lensing region for each profile for specific values of $\lambda$; in this case for $\lambda=1$. For BEC and MSFDM such regions are not unique (because of the zeros of the Bessel functions) and there are several intervals where the jacobian is negative; here are only noted the first intervals of these functions. }\label{strongLENS}
\end{table}

\section{Conclusions}
In this work, the analytical imaging formation process for dark matter halos has been performed. The goal is to study the optical information provided by the lens mapping, to later draw the corresponding images, which emerges naturally by using the $\mathbf{X}$ field, encoded into Eq. (\ref{Xreal}). All this for understanding the physical processes that lies in the core of some types of galaxies.\\
The motivation behind this procedure, is to obtain the analytic equations for the optical mapping from the source plane to the lens plane, and translate directly such information into the images generated on the latter plane. This was approached by obtaining the deflection angle and applying the criterions for identifying where these systems have strong lensing (using the jacobian of the mapping), in such a way that the differences between them can be spotted in the plots.\\
Specifically, these differences begin with the characterization of the caustics on each case, because this generalizes the perfect alignment condition between the luminous source and the lens system \cite{Herrera:strong}, showing that the presence of Einstein rings occurs when the central branch of the caustic is in contact with the source plane (as seen in table \ref{centralXbranch}); if the central branch of the caustic doesn't touch the lens plane, there is no formation of Einstein rings. Hence, the presence of Einstein rings is directly related with a physical quantity of the system: the caustic.\\
On the other hand, we saw that the positions of the images produced in each configuration, give clues for identifying each profile. For example, the MSFDM profile presents a unique characteristic in generating Einstein rings, because for certain fixed $\lambda$'s it generates multiple rings (three in our example, as seen in plots \ref{CloseTriple}), that observationally could be seen as an unique thick ring; we believe that this could be a trail to search differences in the behavior of DM halos.\\
The procedure proves to be successful for determining how the fringes that correspond to one dimensional sources (galaxies, stars or other cosmic objects), distort into the corresponding arclets that usually appear in several astronomic images, and simultaneously in the analysis of the Einstein rings produced in each case. We observed that the image deformations are related directly with the positions and sizes of the sources, and with the deflection angles generated by each profile. In turn, these conditions define the domain where the lenses generate images as functions of the $x_{s}$ (the solutions of the lens equation), because such data measures the behavior of the deflection angle $\alpha_{*}(x)$ in each case.\\
The above, allow us to compare how intense (from the point of view of lenses) are these profiles, because the dynamic conditions from one to another change, depending on the physical conditions that define them. Therefore, the ultimate goal for studying such process is to address the problems of the $\Lambda\textrm{CDM}$ on a small scale, because the NFW profile has problems in this region and alternative models must be tested there. This means to have at hand an observational tool for testing the possible wave nature of the DM or the species diversity, basing the calculations on analysis of the soliton region, because it seems that the observational properties of this zone, is established by means of an universal invariant $\mu_{DM}$ (with the correct values of $r_{s}$ and $\rho_{s}$).\\
Finally, as a future goal based on this work, we have find out that it is possible to apply the method for different zones of interest simultaneously (the disk, the bulge, the SFDM halo and the central supermassive black hole SMBH) \cite{Ana:conseq}, for testing if the images can detect the possible presence of gravitational alterations as, for example, gravitational waves.

\section*{Acknowledgments}
This work was partially supported by CONACyT M\'exico under grants A1-S-8742, 304001, 376127, 240512, FORDECYT-PRONACES grant No. 490769 and I0101/131/07 C-234/07 of the Instituto Avanzado de Cosmolog\'ia (IAC) collaboration (http://www.iac.edu.mx/) and from Sistema Nacional de Investigadores (SNI).\\

\appendix
\section{Normalized functions}\label{AppSur}
Here are listed the corresponding functions to obtain the normalized surface mass density $\Sigma_{*}(x)$ and normalized deflection angle $\alpha_{*}(x)$ for each case.\\
For the MSFDM profile, the $F(x)_{MS}$ function for the surface mass density is given by
\begin{equation}\label{MStateSigma}
    F(x)_{MS}= \pi\left(J_{0}(2x)+\frac{\pi}{2} [J_{1}(2x)\mathbf{H}_{0}(2x)-J_{0}(2x)\mathbf{H}_{1}(2x)]\right),
  \end{equation}
and the $g(x)_{MS}$ function is
\begin{equation}
\begin{array}{l}
\displaystyle g(x)_{MS}=\pi^{2} x\bigg[J_{1}(2x)\\[2ex]
\displaystyle +2\pi \bigg(x[J_{1}(2x)\mathbf{H}_{0}(2x)-J_{0}(2x)\mathbf{H}_{1}(2x)]  \\[2ex]
\displaystyle  -\frac{1}{\pi}[ 2x^{2}\,\,{}_{1}F_{2}\left(1;2,2;-x^{2}\right)J_{1}(2x) \\[2ex]
\displaystyle -x^{3}\,\,{}_{1}F_{2}\left(1;2,3;-x^{2}\right)J_{0}(2x)] \bigg) \bigg],
\end{array}
\end{equation}
where $J_{\nu}$ are the Bessel functions of the first kind of $\nu$-th order, $\mathbf{H}_{\mu}$ are the Struve functions of $\mu$-th
 order and ${}_{p}F_{q}$ is the generalized hypergeometric function \cite{Grads:table,Abramowitz:handbook}.\\
The $F(x)_{NFW}$ function for the NFW profile is given by \cite{Golse:ellip}
\begin{equation}\label{NFWfunction}
F(x)_{NFW}=\left\{ \begin{array}{lcl}
                   \displaystyle \frac{1}{x^{2}-1}\left( 1-\frac{1}{\sqrt{1-x^{2}}}\arccos\!\textrm{h}\frac{1}{x}\right) & {} & (x<1),\\[3ex]
                   \displaystyle \frac{1}{3} & {} & (x=1),\\[3ex]
                   \displaystyle \frac{1}{x^{2}-1}\left( 1-\frac{1}{\sqrt{x^{2}-1}}\arccos\frac{1}{x}\right) & {} & (x>1),
                   \end{array}\right.
\end{equation}
and the $g(x)_{NFW}$ for the deflection angle is
\begin{equation}\label{NFWfunction}
g(x)_{NFW}=\left\{ \begin{array}{lccl}
                   \displaystyle \ln\frac{x}{2}+\frac{1}{\sqrt{1-x^{2}}}\arccos\!\textrm{h}\frac{1}{x} & {} & {} & (x<1),\\[3ex]
                   \displaystyle 1+\ln\frac{1}{2} & {} & {} & (x=1),\\[3ex]
                   \displaystyle \ln\frac{x}{2}+\frac{1}{\sqrt{x^{2}-1}}\arccos\frac{1}{x} & {} & {} & (x>1).
                   \end{array}\right.
\end{equation}
Finally, the values of the constants that appear in the Wave DM profile are obtained through the double factorial defined as \cite{Grads:table}
\begin{displaymath}
\begin{array}{l}
\displaystyle (2n+1)!!=1\cdot3\cdot 5\cdots n,\\[2ex]
\displaystyle (2n)!!= 2\cdot4\cdot6\cdots 2n,
\end{array}
\end{displaymath}
for odd or even cases, respectively. Therefore, we have that $\pi 11!!/(2^{6}7!)\approx 0.1012$ and $\pi13!!/(2^{7}7!)\approx 0.658077$.\\

\end{document}